\documentclass{article}

\usepackage[main,final]{neurips_2026}

\usepackage[utf8]{inputenc} 
\usepackage[T1]{fontenc}    
\usepackage{tikz}
\usetikzlibrary{arrows.meta,positioning,fit,backgrounds,calc}
\usepackage{algorithm}
\usepackage{algpseudocode}
\usepackage{hyperref}       
\usepackage{url}            
\usepackage{booktabs}       
\usepackage{longtable}
\usepackage{amsfonts}       
\usepackage{nicefrac}       
\usepackage{microtype}      
\usepackage{xcolor}         
\usepackage{xspace}
\usepackage{booktabs}   
\usepackage{tabularx}   
\usepackage{amsmath}
\usepackage{graphicx}
\usepackage{multirow}
\usepackage[table]{xcolor}
\usepackage{wrapfig}
\usepackage{needspace}  
\usepackage{makecell}
\usepackage{etoc}          

\newcommand{\crchange}[1]{\textcolor{red}{#1}}

\newcommand{\tool}{$\textsc{BioBlobs}$\xspace}
\newcommand{\mean}{$\textsc{Mean Pooling}$\xspace}

\newcommand{\atten}{$\textsc{Attention Pooling}$\xspace}

\title{BioBlobs: Unsupervised Discovery of Functional Substructures}

\author{%
  Xin Wang$^{1,2}$, Kaiwen Shi$^{1}$, Carlos Oliver$^{1}$ \\
  $^{1}$Vanderbilt University, Nashville, Tennessee, USA \\
  $^{2}$Yale University, New Haven, Connecticut, USA
}

\begin{document}
\etocdepthtag.toc{main}

\maketitle

\begin{abstract}
Protein function is often localized to cohesive substructures, such as catalytic triads, binding pockets, and structural motifs, that occupy only a small fraction of a protein's residues. Yet existing predictive pipelines typically aggregate residue-level embeddings and do not explicitly model or infer these functional substructures, leaving the central biological question unanswered: \emph{which substructure of a protein is responsible for its function?} We introduce \tool, an encoder-agnostic, end-to-end differentiable framework that compresses a protein into a small set of cohesive substructures (\emph{blobs}) and predicts function from these blobs alone, so that each blob corresponds to a candidate functional region. Across diverse protein function prediction tasks and both sequence-only and structure-aware protein language models, \tool is competitive with strong baselines while operating on only a small fraction of residues. Trained only on protein function labels, \tool recovers expert-curated functional sites in InterPro and experimentally annotated catalytic sites in the Mechanism and Catalytic Site Atlas, demonstrating unsupervised functional substructure discovery and opening a path to large-scale functional site discovery in unannotated proteins. The discovered \emph{blobs} adapt their spatial scale to the task, ranging from local catalytic sites to entire structural domains. Code is available at \url{https://github.com/OliverLaboratory/BioBlobs}.
\end{abstract}

\section{Introduction}
\label{sec:intro}
\vspace{-1ex}
Proteins are macromolecules that drive fundamental biological processes across the tree of life.
A protein’s ability to carry out its function is rooted in its specific three-dimensional arrangement—its fold.
For a long time, a key obstacle to assigning function was the difficulty of determining the structure for a given sequence.
This bottleneck has been alleviated by accurate structure predictors such as the AlphaFold \mbox{\citep{af2}} and ESM \mbox{\citep{esm}} families, yielding databases with hundreds of millions of predicted structures.
With structures now broadly available, the central challenge shifts to learning representations that support downstream tasks like function prediction and protein design \mbox{\citep{Watson2023,dauparas2022robust}}.
Protein representation learning (PRL) \mbox{\citep{prl}} addresses this by training neural encoders that map a protein's sequence and/or structure into per-residue embeddings that capture its structural and biophysical properties. These encoders are either trained end-to-end for a specific task, or pre-trained via self-supervision on large unlabelled corpora and reused as general-purpose backbones \mbox{\citep{esm,gearnet,pst}}.


A central goal in studying protein function is to identify which substructures of a protein are responsible for its function. These functional substructures are typically cohesive, modular units: catalytic triads (Ser--His--Asp), Rossmann-like nucleotide-binding cores, and P-loop (Walker A/B) NTPase sites are paradigmatic examples \citep{dodson1998catalytic,rossmann1974chemical,leipe2003evolution}. Most existing PRL pipelines, however, do not model proteins at this substructure level. They use these embeddings in one of two ways. Residue-level approaches train a predictor to label each position as binding-site, catalytic, or otherwise functional \mbox{\citep{scannet,masif}}. This is conceptually direct but limited by the scarcity of residue-level annotations, which are expensive to obtain at scale. Moreover, a single model typically predicts only one type of site, without capturing the diverse functional roles that different substructures carry. Protein-level approaches, used for tasks such as enzyme function or fold classification \mbox{\citep{deepfri}}, instead aggregate residue embeddings into a single vector through pooling, discarding the local spatial organization that defines functional substructures. Because these substructures are small relative to the whole protein, their signals are diluted by surrounding non-functional residues, and the resulting representation offers little insight into which substructure underpins a given prediction.

We propose \tool, an end-to-end differentiable framework that compresses a protein into a small set of cohesive substructures, called \emph{blobs}, and predicts protein function from these blobs alone. A neural blob partitioner produces blobs via a differentiable seed-and-expand procedure, regularized by Hoyer-Square sparsity to keep blobs compact. A multiple-instance learning (MIL) predictor then aggregates blob-level evidence into a protein-level prediction, with blob-wise attention weights doubling as interpretable importance scores. Our contributions are as follows:
{\setlength{\leftmargini}{2.5em}
\begin{itemize}
    \item \textbf{The \tool framework.} We argue that the protein function is best modeled at the substructure level and provide \tool as an encoder-agnostic, end-to-end differentiable framework which makes protein function predictions directly interpretable as localized regions.

    \item \textbf{Strong empirical performance on protein function prediction.} Across diverse protein function tasks from ProteinShake \citep{Kucera2023ProteinShake} and VenusX \citep{tan2025venusx} and protein encoders, \tool is competitive with strong baselines while operating on only a small fraction of residues.

    \item \textbf{Alignment with known functional units.} \tool localizes annotated functional sites in VenusX and recovers experimentally annotated catalytic sites in the M-CSA database \citep{ribeiro_mechanism_2018} from protein-level function supervision alone, opening a path to large-scale functional site discovery in the unannotated proteome. An interactive viewer at \mbox{\url{https://oliverlaboratory.com/resources/bioblobs/}} lets readers compare these blobs with the annotated catalytic residues of \mbox{M-CSA} enzymes.

    \item \textbf{The natural granularity of functional units.} Ablations on diverse function prediction tasks show that higher blob coverage helps structural classification, whose labels span entire domains, but not enzyme classification, which depends on local catalytic sites.

\end{itemize}}

\vspace{-1ex}
\section{\tool Framework}
\label{sec:method}
\vspace{-1ex}

Proteins carry out their functions through substructures that differ in size, shape and the residues involved. A catalytic site may consist of a few residues, whereas a structural domain can span hundreds, and the same kind of site can still differ from one protein to another. When these substructures are fixed in advance, for example as $k$-hop neighborhoods or voxel grids, a model can cut a functional assembly apart, much like trying to understand a bicycle by cutting it into uniformly sized cubes. \tool{} instead learns its blobs from function, guided by three properties. Blobs should be \textbf{spatially compact}, because residues that act together, such as those of a catalytic triad, lie close in space even when they are far apart in sequence. Blobs should be \textbf{selective}, because function typically rests on a small part of the protein. Blobs should be \textbf{learnable from protein-level labels} alone, which requires making the discrete choice of where to place each blob compatible with gradient-based training. The design choices below follow from these properties.

\begin{figure}[ht]
    \centering
    \includegraphics[width=1\linewidth]{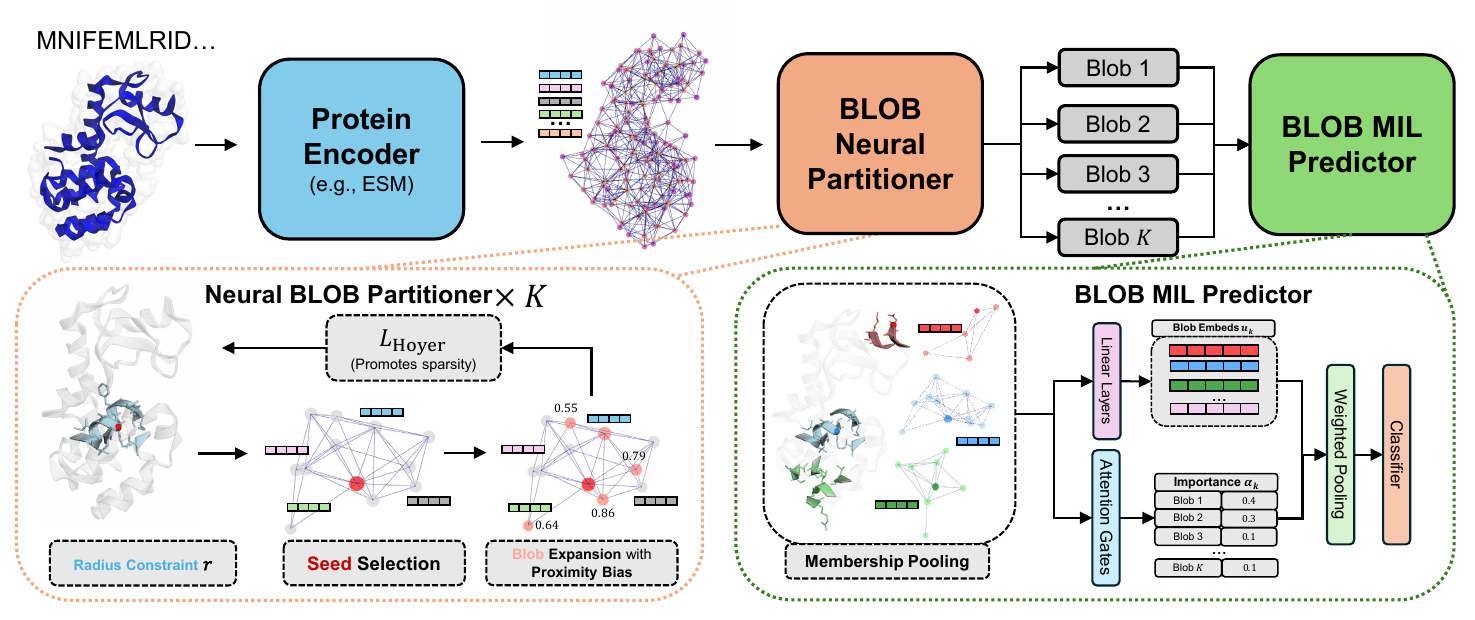}
    \caption{\textbf{The \tool{} pipeline.} A protein encoder maps the protein to residue embeddings. The neural blob partitioner selects $K$ seed residues and expands each into a blob within radius $r$, and the Hoyer-Square penalty keeps the blobs compact. The MIL predictor pools the residue embeddings of each blob, weights the blobs by their importance scores $\alpha_k$ and classifies the protein.}
    \label{fig:bioblobs_pipeline}
\end{figure}

The \tool framework dynamically partitions a protein into a set of cohesive substructures, termed \emph{blobs}, and classifies the protein through attention-based aggregation over them. As shown in Figure~\ref{fig:bioblobs_pipeline}, the pipeline consists of three modules: (1) a protein encoder that produces residue-level embeddings, (2) a neural blob partitioner that proposes blob regions and computes blob embeddings, and (3) a multiple-instance learning (MIL) predictor that aggregates blob-level embeddings into a protein-level prediction. Figure~\ref{fig:pipeline_detailed} and Algorithm~\ref{alg:bioblobs} in Appendix~\ref{appendix:pipeline} detail each step.
\vspace{-1ex}
\subsection{Protein Encoder}\label{sec:encoder}
\tool{} takes as input a protein with $N$ residues, given by its amino acid sequence over the standard 20-letter alphabet $\mathcal{A}$ and its 3D structure, which we represent by the C$_\alpha$ coordinates $\mathbf{X} = [\mathbf{x}_1, \dots, \mathbf{x}_N] \in \mathbb{R}^{N \times 3}$. A protein encoder maps the sequence, optionally incorporating structural features, to residue-level embeddings $\mathbf{Z} = [\mathbf{z}_1, \dots, \mathbf{z}_N] \in \mathbb{R}^{N \times D}$, which can encode context from the whole protein. \tool{} is agnostic to the choice of encoder. The neural blob partitioner uses the coordinates to enforce spatial locality.

\vspace{-1ex}
\subsection{Neural Blob Partitioner}\label{sec:partitioner}
Given residue embeddings $\mathbf{Z} \in \mathbb{R}^{N \times D}$ and coordinates $\mathbf{X} \in \mathbb{R}^{N \times 3}$, the partitioner discovers up to $K$ local \emph{blobs} for the downstream classifier in two steps. \textbf{Seed selection} identifies $K$ functionally informative anchor residues through a learned, differentiable selection procedure. \textbf{Blob expansion} grows each seed into a soft, spatially local substructure by combining learned semantic compatibility with a distance-based proximity prior over residues within radius $r$ of the seed.

\vspace{-1ex}
\paragraph{Seed selection.}
Seeds act as anchors, and the partitioner learns which residues, such as catalytic residues, are worth building a blob around. A seed scoring network $f_{\text{seed}} : \mathbb{R}^D \to \mathbb{R}$, implemented as a one-hidden-layer MLP with hidden width $d_h$, assigns a scalar score $s_i$ to each residue. To anchor distinct regions of the protein rather than concentrating on a single high-scoring residue, we select $K$ seeds sequentially without replacement: at each step $k$, previously selected residues are masked, and the next seed $v_{\text{seed}}^{(k)}$ is selected as the argmax of a temperature-scaled softmax, $\mathbf{w}^{(k)} = \mathrm{softmax}(\tilde{\mathbf{s}}^{(k)} / \tau_{\text{seed}})$, where $\tilde{\mathbf{s}}^{(k)}$ denotes the masked logits at step $k$. The selected residue comes with the seed embedding $\mathbf{z}_{\text{seed}}^{(k)}$ and its coordinate $\mathbf{x}_{\text{seed}}^{(k)} = \mathbf{x}_{v_{\text{seed}}^{(k)}}$. Since this is a discrete process, we pass gradients to the seed scores with the straight-through estimator \citep{bengio2013estimating, jang2016categorical}, so the function loss can teach the scorer where to anchor.
\vspace{-1ex}
\paragraph{Blob expansion.}
Each seed $k$ expands into a blob by absorbing residues in a spatially local neighborhood. A blob should hold the residues that lie near its seed and, together with it, form a candidate functional site that helps the prediction. The radius $r$ enforces spatial locality, restricting each blob to a compact region of the structure rather than a globally distributed set of residues. We form a candidate set from all residues within radius $r$ of the seed's $\text{C}_\alpha$ coordinate, \mbox{$\mathcal{C}^{(k)} = \{i : \|\mathbf{x}_i - \mathbf{x}_{\text{seed}}^{(k)}\|_2 \leq r\}$}.
For each candidate $i \in \mathcal{C}^{(k)}$, membership is determined by two complementary signals: a \emph{learned attention score} $a_{ki}$ that captures semantic compatibility between the seed and the residue, and a \emph{proximity bias} $\rho_{ki}$ that decays linearly with Euclidean distance, favoring residues close to the seed:
\begin{equation}
    a_{ki} = \frac{(\mathbf{W}_q\, \mathbf{z}_{\text{seed}}^{(k)})^\top (\mathbf{W}_{\text{key}}\, \mathbf{z}_i)}{\sqrt{d_a}}, \qquad \rho_{ki} = \max\!\left(0,\; 1 - \frac{\|\mathbf{x}_i - \mathbf{x}_{\text{seed}}^{(k)}\|_2}{r}\right),
    \label{eq:attn_prox}
\end{equation}
where $\mathbf{W}_q, \mathbf{W}_{\text{key}} \in \mathbb{R}^{d_a \times D}$ are learnable projections. The combined logit is passed through a sigmoid to yield a soft membership score:
\begin{equation}
    m_{ki} = \sigma\!\left(\ell_{ki} / \tau_{\text{mem}}\right), \qquad \ell_{ki} = a_{ki} + \lambda_\rho\, \rho_{ki}, \qquad i \in \mathcal{C}^{(k)},
    \label{eq:membership}
\end{equation}
where $\lambda_\rho$ weights the proximity bias and $\tau_{\text{mem}}$ is annealed on the same schedule as $\tau_{\text{seed}}$, as detailed in Appendix~\ref{appendix:pipeline}. Restricting candidates to radius $r$ prevents semantically similar but spatially distant residues from being merged into the same blob. Within the radius, a blob can gather residues that are adjacent in space but distant in sequence, and the attention score can leave out nearby residues that do not belong to the same blob. The seed residue is hard-clamped to $m_{k,\, v_{\text{seed}}^{(k)}} = 1$ to guarantee that every blob contains at least its seed. The full soft assignment matrix $\mathbf{M} \in [0,1]^{K \times N}$ records the blob membership of each residue in each blob. Entries outside each candidate set are zero, and blobs may overlap because memberships are computed independently for each blob, so a residue can belong to more than one blob.

\vspace{-1ex}
\subsection{Multiple Instance Learning (MIL) Predictor}\label{sec:mil}
The MIL predictor makes a protein-level prediction from the blobs and learns which blobs the prediction relies on. Given the blob assignment matrix $\mathbf{M}$, the embedding for blob $k$ is the membership-weighted mean of residue features:
\begin{equation}
    \mathbf{b}_k = \frac{\sum_{i=1}^{N} m_{ki}\, \mathbf{z}_i}{\sum_{i=1}^{N} m_{ki} + \varepsilon},
    \label{eq:blob_embedding}
\end{equation}
where $\varepsilon$ is a small constant for numerical stability. The matrix $\mathbf{B} = [\mathbf{b}_1, \dots, \mathbf{b}_K] \in \mathbb{R}^{K \times D}$ collects the blob embeddings of a protein. Blob embeddings can optionally be quantized with residual vector quantization (RVQ)~\citep{zeghidour2021soundstream}, which writes each normalized blob embedding as a sum of $Q$ code vectors, one from each of $Q$ learned codebooks $\{\boldsymbol{\mu}_{q,j}\}_{j=1}^{V}$ shared across proteins. Starting from $\boldsymbol{\delta}_{k,0}=\mathbf{b}_k/\|\mathbf{b}_k\|_2$, each stage $q=1,\dots,Q$ selects the code vector nearest to the residual left by the preceding stages,
\begin{equation}
    j_{k,q}=\arg\min_{j}\,\|\boldsymbol{\delta}_{k,q-1}-\boldsymbol{\mu}_{q,j}\|_2^2,\qquad
    \boldsymbol{\delta}_{k,q}=\boldsymbol{\delta}_{k,q-1}-\boldsymbol{\mu}_{q,j_{k,q}},\qquad
    \widehat{\mathbf{b}}_k=\sum_{q=1}^{Q}\boldsymbol{\mu}_{q,j_{k,q}},
    \label{eq:rvq}
\end{equation}
and $\widehat{\mathbf{b}}_k$ replaces $\mathbf{b}_k$ in the MIL predictor below. RVQ is found to help functional site localization, as shown in Appendix~\ref{appendix:rvq}.
Following the attention-based MIL framework~\citep{ilse2018attention}, we treat the protein as a bag and its $K$ blobs as instances. Some instances may carry the key functional signal, for example the blob that covers an active site. A learned feature transform first maps each blob embedding,
$\mathbf{u}_k = \mathrm{ReLU}(\mathbf{W}_\phi\, \mathbf{b}_k + \mathbf{c}_\phi),$
where $k = 1, \dots, K$, and $\mathbf{W}_\phi$ and $\mathbf{c}_\phi$ are learnable. An attention gate then computes a scalar importance score for each blob,
\begin{equation}
    \alpha_k = \frac{\exp(\mathbf{w}_a^\top \mathbf{b}_k + c_a)}{\sum_{j=1}^{K} \exp(\mathbf{w}_a^\top \mathbf{b}_j + c_a)},
    \label{eq:mil_attention}
\end{equation}
where $\mathbf{w}_a$ and $c_a$ are learnable, and the softmax is taken over valid blobs only. The weights $\boldsymbol{\alpha} = [\alpha_1, \dots, \alpha_K]$ serve as interpretable importance scores indicating each blob's contribution to the prediction. Decoupling the feature transform $\mathbf{u}_k$ from the importance score $\alpha_k$ lets the model learn \emph{what} each blob contributes separately from \emph{how much} it contributes. The protein-level representation is the attention-weighted sum $\mathbf{z}_{\text{bag}} = \sum_{k=1}^{K} \alpha_k\, \mathbf{u}_k$, which is passed to an MLP prediction head for classification. After training, each residue can be scored by its maximum membership over blobs, $\max_k m_{ki}$, or by its attention-weighted membership, $\sum_k \alpha_k m_{ki}$. We use these scores to evaluate functional site localization and visualize discrete blobs for interpretability.

\vspace{-1ex}
\subsection{Training Objectives}\label{sec:objectives}
If we trained with only the task loss (e.g., cross-entropy for multi-class classification), the partitioner tends to absorb many of the candidates in each seed's spherical neighborhood, contradicting our goal of \emph{identifying a minimal set of residues sufficient for function prediction}. We therefore regularize the soft membership matrix $\mathbf{M}$ with \textbf{Hoyer's scale-invariant sparseness measure}~\citep{hoyer2004non}, originally proposed for non-negative matrix factorization with sparse, localized basis columns and later adapted by DeepHoyer~\citep{yang2019deephoyer} for sparse neural networks. Hoyer fits our setting because it depends only on the shape of each membership vector rather than its total mass, and is evaluated independently per blob, allowing different blobs to settle at different compact sizes. This is appropriate when functional substructures range in size, from 3-residue catalytic triads to roughly 30-residue Rossmann cores.
For each valid blob $k$, let $\mathbf{m}_k \in [0,1]^{n_k}$ be the membership vector over its $n_k = |\mathcal{C}^{(k)}|$ candidates. We define the Hoyer-Square value as
\begin{equation}
    \mathrm{HS}_k \;=\; \frac{1}{n_k}\!\left(\frac{\lVert \mathbf{m}_k \rVert_1}{\lVert \mathbf{m}_k \rVert_2}\right)^{\!2} \;\in\; \left[\tfrac{1}{n_k},\; 1\right],
    \label{eq:hoyer}
\end{equation}
which equals $1$ for uniform $\mathbf{m}_k$ and $1/n_k$ for one-hot $\mathbf{m}_k$. The product $n_k \cdot \mathrm{HS}_k$ can be read as the effective number of residues to which blob $k$ commits. The training objective is
\begin{equation}
    \mathcal{L} \;=\; \mathcal{L}_{\text{task}} \;+\; \lambda_{\text{H}}\, \mathcal{L}_{\text{Hoyer}}, \qquad \mathcal{L}_{\text{Hoyer}} \;=\; \frac{1}{|\mathcal{V}_{\text{blob}}|}\sum_{k \in \mathcal{V}_{\text{blob}}} \mathrm{HS}_k,
    \label{eq:objective}
\end{equation}
where $\mathcal{V}_{\text{blob}}$ denotes the set of validly-seeded blobs in the batch, i.e., those whose seeds correspond to real residues rather than padded positions in batched inputs of varying length. When RVQ is enabled, the objective also includes its commitment penalty, defined in Appendix~\ref{appendix:rvq}. The codebooks are updated by exponential moving averages.

\vspace{-1ex}
\section{Experiments}
\vspace{-1ex}
\label{sec:experiments}
We evaluate \tool{} on two tasks. Protein function prediction tests whether compact blob representations support accurate protein-level predictions, and functional substructure discovery tests whether blobs learned from protein labels alone localize annotated functional sites. We describe the benchmark datasets, baselines, and evaluation framework used throughout our experiments.
\vspace{-1ex}
\subsection{Datasets}
\label{sec:dataset}
We benchmark \tool on two suites, detailed in Appendix~\ref{appendix:datasets}. \textbf{ProteinShake}~\citep{Kucera2023ProteinShake} provides five protein-level prediction tasks covering molecular function (\texttt{GO-MF}), reaction catalysis (\texttt{EC-L3}), structural classification (\texttt{SCOP-FAM}, \texttt{SCOP-SF}) and domain families (\texttt{Pfam}), each evaluated under a sequence split and a structure split at a 0.7 similarity threshold, which assign whole clusters of similar sequences or structures to disjoint splits. \textbf{VenusX}~\citep{tan2025venusx} provides four datasets in which each protein is labeled with the InterPro entry of an annotated functional site, covering catalytic active sites (\texttt{Act}), ligand-binding interfaces (\texttt{BindI}), sites under strong evolutionary pressure (\texttt{Evo}) and conserved motifs (\texttt{Motif}), with the provided Mixed-Family (MF-50) splits. All methods take the full protein chain as input, and the annotated fragment serves only as the evaluation target.

\textbf{M-CSA}~\citep{ribeiro_mechanism_2018} curates catalytic residues individually from the literature for enzymes with known mechanisms. Enzyme Commission (EC) numbers describe the reaction an enzyme catalyzes at four levels of increasing detail, and by the third level they specify features such as the bond acted on, the group transferred or the acceptor (e.g., EC~2.1.1, methyltransferases). For this evaluation, we train on the third-level EC classes of SwissProt enzymes~\citep{uniprot_consortium_uniprot_2025} from EC-Bench~\citep{davoudi_ec-bench_2026}, with AlphaFold v6 structures~\citep{varadi_alphafold_2024}, and produce train and validation partitions via \textit{MMseqs2}~\citep{steinegger_mmseqs2_2017} at a 0.3 sequence-identity threshold. We evaluate on M-CSA enzymes, and the 74 that share no 30\%-identity cluster with the training set form the held-out set.
\vspace{-1ex}
\subsection{Baselines and Evaluation Framework}
\label{sec:baselines_evaluation}
We use frozen residue embeddings from ESM2-T30~\citep{esm}, a sequence-only protein language model, and SaProt-650M~\citep{su_saprot_2023}, which adds Foldseek structure tokens. For function prediction, we compare \tool{} with mean and attention pooling~\citep{ilse2018attention} on both suites, and on ProteinShake with three graph neural networks, GCN~\citep{kipf2016semi}, GIN~\citep{xu2018powerful} and GAT~\citep{velivckovic2017graph}, and three graph-pooling methods, DiffPool~\citep{ying2018hierarchical}, TopK~\citep{gao2019graph} and SAGPool~\citep{lee2019self}. For functional site localization, the baselines rank residues by the pooling classifiers' attention weights or gradient-based attributions such as integrated gradients~\citep{sundararajan2017axiomatic}, with geometric burial and random scores as controls. On M-CSA, we also compare with ProtDETR~\citep{yang2025interpretable}, ProteInfer~\citep{sanderson_proteinfer_2023} and DeepFRI~\citep{deepfri}, published EC predictors that provide residue-level saliency, and with two label-free signals from ESM2: its attention maps and a sparse autoencoder trained on its embeddings~\citep{bussmann2024batchtopk}. We report test macro-F1 for multi-class tasks and $F_{\max}$ for multi-label tasks, and evaluate localization against annotated residues with AUROC, AUPRC and, on M-CSA, AUPRC enrichment over the fraction of annotated residues. ProteinShake and VenusX results are means and sample standard deviations over three seeds. Appendices~\ref{appendix:implementation} and~\ref{appendix:localization} give the training protocol, baselines and metric definitions.




\vspace{-1ex}
\section{Results}
\vspace{-1ex}

In this section, we answer the following questions. \textbf{Performance:} Does \tool{} match or exceed pooling baselines on protein function prediction? \textbf{Recovery:} Do the blobs recover human-annotated functional sites? \textbf{Granularity:} Do the discovered substructures adapt their granularity to the task while remaining compact and informative for prediction? \textbf{Cost:} How much time and memory does blob construction add?

\vspace{-1ex}
\subsection{\tool{} is competitive with pooling baselines on protein function prediction}
\label{sec:results-function}

\tool{} predicts function from the residues in its blobs alone, yet it is competitive with both pooling and graph-based baselines. With ESM2-T30, the blobs hold about a quarter of a protein's residues or fewer on four of the five tasks. Against \mean{} and \atten{}, \tool{} is best in six of the ten dataset and split cells, and against the six graph-based baselines in seven. Where another method leads, the margin is at most $0.023$. Distilling a protein into a small subset of residues thus need not reduce prediction accuracy, consistent with the view that protein function is carried by substructures. The advantage over pooling extends to the structure-aware SaProt-650M, where \tool{} outperforms both pooling baselines on every \texttt{Pfam} and \texttt{SCOP-FAM} cell. Table~\ref{tab:venusx_classification} in Appendix~\ref{appendix:localization} reports the VenusX classification results, which show similar patterns.

\begin{table}[ht]
\centering
\caption{\textbf{Protein function prediction on ProteinShake.} Bold marks the best mean for each encoder, split and dataset.}
\label{tab:proteinshake}
\scriptsize
\setlength{\tabcolsep}{3pt}
\resizebox{\textwidth}{!}{%
\begingroup\color{black}\renewcommand{\crchange}[1]{#1}%
\begin{tabular}{llccccc}
\toprule
Encoder & \crchange{Method} & \texttt{EC-L3} & \texttt{Pfam} & \texttt{GO-MF} & \texttt{SCOP-FAM} & \texttt{SCOP-SF} \\
\midrule
\multicolumn{7}{l}{\textbf{Sequence split}} \\
ESM2-T30 & Mean pooling & $0.562\pm0.009$ & $0.381\pm0.004$ & $0.743\pm0.002$ & $0.340\pm0.011$ & $0.519\pm0.012$ \\
 & Attention pooling & \crchange{$0.594\pm0.004$} & \crchange{$0.398\pm0.015$} & \crchange{$\mathbf{0.756\pm0.005}$} & \crchange{$0.361\pm0.005$} & \crchange{$\mathbf{0.534\pm0.012}$} \\
 & \crchange{GCN} & \crchange{$0.502\pm0.082$} & \crchange{$0.366\pm0.016$} & \crchange{$0.746\pm0.010$} & \crchange{$0.338\pm0.008$} & \crchange{$0.495\pm0.024$} \\
 & \crchange{GIN} & \crchange{$0.525\pm0.056$} & \crchange{$0.390\pm0.030$} & \crchange{$0.744\pm0.010$} & \crchange{$0.336\pm0.007$} & \crchange{$0.509\pm0.031$} \\
 & \crchange{GAT} & \crchange{$0.532\pm0.061$} & \crchange{$0.391\pm0.013$} & \crchange{$0.749\pm0.011$} & \crchange{$0.357\pm0.018$} & \crchange{$0.524\pm0.017$} \\
 & \crchange{DiffPool} & \crchange{$0.572\pm0.065$} & \crchange{$0.401\pm0.006$} & \crchange{$0.755\pm0.001$} & \crchange{$0.373\pm0.017$} & \crchange{$0.512\pm0.018$} \\
 & \crchange{TopK} & \crchange{$0.563\pm0.047$} & \crchange{$0.373\pm0.014$} & \crchange{$0.751\pm0.005$} & \crchange{$0.343\pm0.027$} & \crchange{$0.500\pm0.013$} \\
 & \crchange{SAGPool} & \crchange{$0.550\pm0.045$} & \crchange{$0.390\pm0.006$} & \crchange{$0.754\pm0.002$} & \crchange{$0.321\pm0.005$} & \crchange{$0.490\pm0.018$} \\
 & \tool & $\mathbf{0.615\pm0.004}$ & $\mathbf{0.402\pm0.015}$ & $0.754\pm0.003$ & $\mathbf{0.377\pm0.003}$ & $0.528\pm0.003$ \\
\cmidrule(lr){1-7}
SaProt-650M & Mean pooling & $0.586\pm0.018$ & $0.362\pm0.006$ & $0.751\pm0.004$ & $0.353\pm0.004$ & $0.543\pm0.022$ \\
 & Attention pooling & \crchange{$\mathbf{0.688\pm0.050}$} & \crchange{$0.400\pm0.013$} & \crchange{$\mathbf{0.763\pm0.004}$} & \crchange{$0.360\pm0.008$} & \crchange{$0.565\pm0.015$} \\
 & \crchange{GCN} & \crchange{$0.587\pm0.053$} & \crchange{$0.358\pm0.001$} & \crchange{$0.747\pm0.005$} & \crchange{$0.353\pm0.012$} & \crchange{$0.543\pm0.008$} \\
 & \crchange{GIN} & \crchange{$0.544\pm0.119$} & \crchange{$0.363\pm0.034$} & \crchange{$0.746\pm0.010$} & \crchange{$0.338\pm0.031$} & \crchange{$0.518\pm0.008$} \\
 & \crchange{GAT} & \crchange{$0.531\pm0.123$} & \crchange{$0.392\pm0.014$} & \crchange{$0.759\pm0.006$} & \crchange{$\mathbf{0.373\pm0.016}$} & \crchange{$\mathbf{0.570\pm0.001}$} \\
 & \crchange{DiffPool} & \crchange{$0.546\pm0.061$} & \crchange{$0.384\pm0.009$} & \crchange{$0.762\pm0.003$} & \crchange{$0.369\pm0.009$} & \crchange{$0.565\pm0.003$} \\
 & \crchange{TopK} & \crchange{$0.509\pm0.042$} & \crchange{$0.397\pm0.003$} & \crchange{$0.751\pm0.005$} & \crchange{$0.332\pm0.005$} & \crchange{$0.542\pm0.018$} \\
 & \crchange{SAGPool} & \crchange{$0.579\pm0.050$} & \crchange{$0.377\pm0.023$} & \crchange{$0.748\pm0.005$} & \crchange{$0.323\pm0.021$} & \crchange{$0.506\pm0.004$} \\
 & \tool & \crchange{$0.647\pm0.047$} & \crchange{$\mathbf{0.403\pm0.012}$} & \crchange{$0.745\pm0.003$} & \crchange{$0.371\pm0.009$} & \crchange{$0.565\pm0.010$} \\
\midrule
\multicolumn{7}{l}{\textbf{Structure split}} \\
ESM2-T30 & Mean pooling & $0.560\pm0.051$ & $0.213\pm0.009$ & $0.643\pm0.013$ & $0.203\pm0.012$ & $0.381\pm0.009$ \\
 & Attention pooling & \crchange{$0.544\pm0.066$} & \crchange{$0.228\pm0.003$} & \crchange{$\mathbf{0.660\pm0.012}$} & \crchange{$0.220\pm0.008$} & \crchange{$0.410\pm0.013$} \\
 & \crchange{GCN} & \crchange{$0.388\pm0.146$} & \crchange{$0.210\pm0.013$} & \crchange{$0.643\pm0.002$} & \crchange{$0.193\pm0.009$} & \crchange{$0.345\pm0.006$} \\
 & \crchange{GIN} & \crchange{$0.464\pm0.041$} & \crchange{$0.228\pm0.005$} & \crchange{$0.615\pm0.014$} & \crchange{$0.189\pm0.011$} & \crchange{$0.361\pm0.027$} \\
 & \crchange{GAT} & \crchange{$0.434\pm0.067$} & \crchange{$0.223\pm0.012$} & \crchange{$0.654\pm0.001$} & \crchange{$0.198\pm0.011$} & \crchange{$0.383\pm0.006$} \\
 & \crchange{DiffPool} & \crchange{$0.525\pm0.125$} & \crchange{$0.221\pm0.020$} & \crchange{$0.655\pm0.011$} & \crchange{$0.223\pm0.016$} & \crchange{$\mathbf{0.420\pm0.011}$} \\
 & \crchange{TopK} & \crchange{$0.508\pm0.118$} & \crchange{$0.214\pm0.022$} & \crchange{$0.635\pm0.005$} & \crchange{$0.189\pm0.016$} & \crchange{$0.359\pm0.011$} \\
 & \crchange{SAGPool} & \crchange{$0.429\pm0.070$} & \crchange{$0.213\pm0.006$} & \crchange{$0.634\pm0.021$} & \crchange{$0.186\pm0.041$} & \crchange{$0.345\pm0.016$} \\
 & \tool & $\mathbf{0.592\pm0.026}$ & $\mathbf{0.232\pm0.009}$ & $0.655\pm0.022$ & $\mathbf{0.256\pm0.010}$ & $0.397\pm0.023$ \\
\cmidrule(lr){1-7}
SaProt-650M & Mean pooling & $0.504\pm0.011$ & $0.200\pm0.017$ & $0.643\pm0.004$ & $0.185\pm0.004$ & $0.385\pm0.015$ \\
 & Attention pooling & \crchange{$\mathbf{0.551\pm0.028}$} & \crchange{$0.216\pm0.008$} & \crchange{$\mathbf{0.673\pm0.013}$} & \crchange{$0.202\pm0.008$} & \crchange{$0.419\pm0.008$} \\
 & \crchange{GCN} & \crchange{$0.536\pm0.027$} & \crchange{$0.192\pm0.008$} & \crchange{$0.622\pm0.008$} & \crchange{$0.184\pm0.018$} & \crchange{$0.399\pm0.006$} \\
 & \crchange{GIN} & \crchange{$0.501\pm0.016$} & \crchange{$0.206\pm0.002$} & \crchange{$0.619\pm0.016$} & \crchange{$0.181\pm0.004$} & \crchange{$0.367\pm0.063$} \\
 & \crchange{GAT} & \crchange{$0.370\pm0.067$} & \crchange{$\mathbf{0.230\pm0.009}$} & \crchange{$0.633\pm0.033$} & \crchange{$0.213\pm0.007$} & \crchange{$0.421\pm0.047$} \\
 & \crchange{DiffPool} & \crchange{$0.481\pm0.064$} & \crchange{$0.219\pm0.005$} & \crchange{$0.661\pm0.005$} & \crchange{$\mathbf{0.224\pm0.027}$} & \crchange{$0.448\pm0.013$} \\
 & \crchange{TopK} & \crchange{$0.347\pm0.094$} & \crchange{$0.196\pm0.016$} & \crchange{$0.661\pm0.017$} & \crchange{$0.199\pm0.020$} & \crchange{$0.343\pm0.055$} \\
 & \crchange{SAGPool} & \crchange{$0.352\pm0.057$} & \crchange{$0.192\pm0.012$} & \crchange{$0.637\pm0.021$} & \crchange{$0.186\pm0.027$} & \crchange{$0.396\pm0.038$} \\
 & \tool & \crchange{$0.544\pm0.054$} & \crchange{$0.217\pm0.005$} & \crchange{$0.645\pm0.006$} & \crchange{$0.216\pm0.007$} & \crchange{$\mathbf{0.453\pm0.018}$} \\
\bottomrule
\end{tabular}\endgroup}
\end{table}

\vspace{-1ex}
\subsection{\tool{} recovers known functional sites without residue-level supervision}
\label{sec:csa_recovery}
While learning to predict protein function, \tool{} also learns where that
function resides. Its blobs align with experimentally annotated functional sites, although
training and model selection use only protein-level labels. We score each residue by its
membership in the blobs. Three metrics compare this residue ranking with the annotated
site, and they differ in which part of the ranking they emphasize. A functional site covers only
a small number of the residues in a full protein chain, so useful localization must rank those
few residues ahead of all others. AUPRC, the area under the precision--recall curve, measures
this directly. A random ranking scores near the small fraction of annotated residues, and AUPRC
approaches one only when the site residues rank first. This random level depends on site
size, so AUPRC is comparable only within a dataset. AUROC, the probability that an annotated
residue ranks above an unannotated one, weighs the whole ranking equally and has the same random
level of $0.5$ for any site size. F1@$L$ evaluates a single predicted site rather than the
ranking. It takes the $L$ top-ranked residues, where $L$ is the number of annotated residues, and
measures the fraction of them in the site. We evaluate on
VenusX, whose four datasets annotate different kinds of sites, and then on catalytic
residues, against the baselines of Section~\ref{sec:baselines_evaluation}.

\vspace{-1ex}
\paragraph{VenusX.}
As shown in Table~\ref{tab:venusx}, \tool{} localizes functional sites better than
the baselines on most datasets and metrics. With SaProt-650M, it leads on every dataset in all
three metrics, and with ESM2-T30 it leads in AUROC on three datasets and in AUPRC on
\texttt{Act} and \texttt{Evo}. Each residue's membership is weighted by the MIL attention of
its blob, $c_i=\sum_k\alpha_k m_{ki}$, so residues rank first when they belong to the blobs
that drive the prediction. The better encoder depends on the site type. SaProt-650M localizes
\texttt{Motif} sites better than ESM2-T30 in all three metrics, ESM2-T30 is better on
\texttt{Act}, and the two are close on \texttt{BindI} and \texttt{Evo}. If the training
proteins come with residue annotations, \tool{} can improve further by adding a
seed-alignment loss that encourages seeds to fall within the annotated sites. With both
encoders, this puts \tool{} above every baseline on all four datasets, and on \texttt{Motif}
it more than doubles the AUPRC of the best baseline. Appendix~\ref{appendix:seed_align}
defines the seed-alignment loss, and Appendix~\ref{appendix:implementation} gives the
experiment details.

\begin{table}[ht]
\centering
\caption{\textbf{Functional site localization on VenusX.} Mean over three seeds. Bold marks the best mean without seed alignment within each encoder.}
\label{tab:venusx}
\footnotesize
\setlength{\tabcolsep}{3pt}
\resizebox{\textwidth}{!}{%
\begin{tabular}{l*{12}{c}}
\toprule
 & \multicolumn{4}{c}{AUROC $\uparrow$} & \multicolumn{4}{c}{AUPRC $\uparrow$} & \multicolumn{4}{c}{F1@$L$ $\uparrow$} \\
\cmidrule(lr){2-5}\cmidrule(lr){6-9}\cmidrule(lr){10-13}
Method & \texttt{Act} & \texttt{BindI} & \texttt{Evo} & \texttt{Motif} & \texttt{Act} & \texttt{BindI} & \texttt{Evo} & \texttt{Motif} & \texttt{Act} & \texttt{BindI} & \texttt{Evo} & \texttt{Motif} \\
\midrule
\multicolumn{13}{l}{\textit{ESM2-T30}} \\
Uniform random & $0.500$ & $0.499$ & $0.498$ & $0.500$ & $0.050$ & $0.057$ & $0.078$ & $0.109$ & $0.000$ & $0.049$ & $0.068$ & $0.109$ \\
Residue burial & $0.795$ & $0.676$ & $0.670$ & $0.577$ & $0.121$ & $0.103$ & $0.120$ & $0.127$ & $0.121$ & $0.105$ & $0.111$ & $0.082$ \\
Integrated gradients & $0.656$ & $0.806$ & $0.659$ & $\mathbf{0.703}$ & $0.083$ & $0.168$ & $0.166$ & $\mathbf{0.302}$ & $0.086$ & $0.174$ & $0.173$ & $\mathbf{0.311}$ \\
Attention pooling & $0.851$ & $0.793$ & $0.751$ & $0.673$ & $0.187$ & $\mathbf{0.246}$ & $0.221$ & $0.215$ & $0.197$ & $\mathbf{0.257}$ & $\mathbf{0.236}$ & $0.236$ \\
\tool{} & $\mathbf{0.886}$ & $\mathbf{0.831}$ & $\mathbf{0.803}$ & $0.662$ & $\mathbf{0.235}$ & $0.207$ & $\mathbf{0.233}$ & $0.228$ & $\mathbf{0.245}$ & $0.197$ & $0.235$ & $0.197$ \\
\cmidrule(lr){1-13}
\tool{} (+ seed align) & $0.873$ & $0.886$ & $0.824$ & $0.902$ & $0.276$ & $0.368$ & $0.290$ & $0.658$ & $0.301$ & $0.376$ & $0.299$ & $0.608$ \\
\midrule
\multicolumn{13}{l}{\textit{SaProt-650M}} \\
Uniform random & $0.500$ & $0.499$ & $0.498$ & $0.500$ & $0.050$ & $0.057$ & $0.078$ & $0.109$ & $0.000$ & $0.049$ & $0.068$ & $0.109$ \\
Residue burial & $0.795$ & $0.676$ & $0.670$ & $0.577$ & $0.121$ & $0.103$ & $0.120$ & $0.127$ & $0.121$ & $0.105$ & $0.111$ & $0.082$ \\
Integrated gradients & $0.591$ & $0.676$ & $0.642$ & $0.693$ & $0.060$ & $0.118$ & $0.126$ & $0.228$ & $0.000$ & $0.110$ & $0.120$ & $0.251$ \\
Attention pooling & $0.767$ & $0.741$ & $0.728$ & $0.678$ & $0.128$ & $0.165$ & $0.196$ & $0.250$ & $0.121$ & $0.188$ & $0.206$ & $0.257$ \\
\tool{} & $\mathbf{0.814}$ & $\mathbf{0.780}$ & $\mathbf{0.776}$ & $\mathbf{0.747}$ & $\mathbf{0.173}$ & $\mathbf{0.211}$ & $\mathbf{0.230}$ & $\mathbf{0.397}$ & $\mathbf{0.177}$ & $\mathbf{0.233}$ & $\mathbf{0.239}$ & $\mathbf{0.395}$ \\
\cmidrule(lr){1-13}
\tool{} (+ seed align) & $0.845$ & $0.845$ & $0.809$ & $0.910$ & $0.199$ & $0.301$ & $0.282$ & $0.692$ & $0.208$ & $0.311$ & $0.290$ & $0.642$ \\
\bottomrule
\end{tabular}}
\end{table}

\vspace{-1ex}
\paragraph{Mechanism and Catalytic Site Atlas (M-CSA).}
We further validate the blobs on catalytic sites, the residues that take part
directly in an enzyme's catalytic mechanism. Enzyme Commission (EC) numbers classify enzymes
by the reactions they catalyze. Catalytic residues are therefore natural evidence for an
enzyme's EC class, and a model that predicts EC classes for the right reasons should locate
them. We train \tool{} with $8$ blobs
and residual vector quantization to predict
the \mbox{third-level} EC classes. A class-balanced cross-entropy loss with effective-number
reweighting \citep{Cui_2019_CVPR} handles the long-tailed classes. We then
compare its blobs and attention with the annotated catalytic residues in M-CSA.

\begin{figure}[t]
    \centering
    \includegraphics[width=\textwidth]{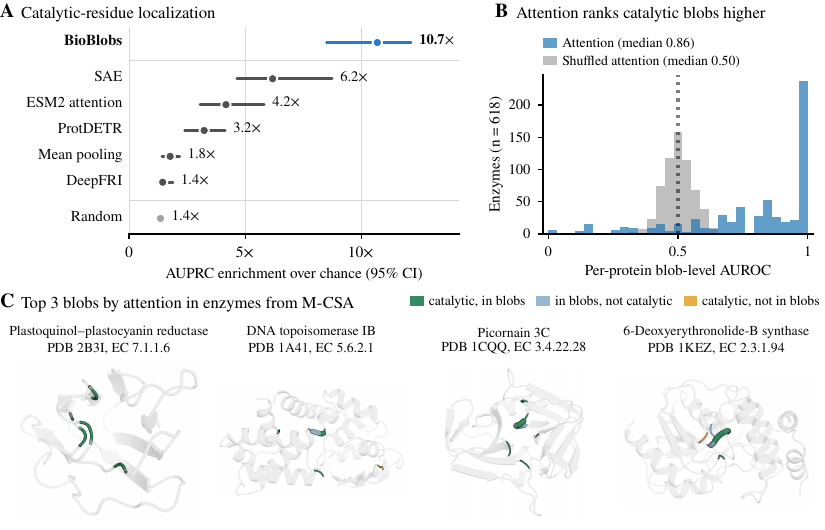}
    \caption{\textbf{\tool{} blobs agree with M-CSA catalytic sites.}
    \textbf{A.} Catalytic-residue localization of 74 held-out M-CSA enzymes.
    \textbf{B.} MIL attention ranks blobs containing an annotated catalytic residue above
    other blobs in the same enzyme, unlike shuffled attention.
    \textbf{C.} Overlap between the three highest-attention blobs and the annotated catalytic
    residues in the four best-matching enzymes.}
    \label{fig:csa_recovery}
\end{figure}

The blobs alone localize catalytic residues better than every baseline, as
Figure~\ref{fig:csa_recovery}A shows. On the 74 held-out enzymes, scoring each residue by its
maximum blob membership $\max_k m_{ki}$, which does not use attention, gives a median AUPRC
enrichment of $10.7\times$ over chance. The strongest baseline, a sparse autoencoder (SAE) trained
without labels on the same ESM2 features, reaches $6.2\times$, and the attention each residue
receives in ESM2 gives $4.2\times$. Enzyme-function predictors that explain their predictions with
attention, activation maps or gradients, such as ProtDETR, ProteInfer and DeepFRI, reach at most
$3.2\times$. The advantage comes from where the blobs form, as they hold 4\% of the residues but
48\% of the catalytic residues.

MIL attention then picks out the blobs that contain catalytic residues, as
Figure~\ref{fig:csa_recovery}B shows. Across M-CSA enzymes, the median per-protein blob-level
AUROC is $0.86$, against $0.50$ for shuffled attention. The held-out enzymes alone give the same
median, so similarity to the training set does not explain this ranking. Figure~\ref{fig:csa_recovery}C shows the four
best-matching enzymes, in which the three highest-attention blobs cover most catalytic residues. Earlier work, such as DeepFRI~\citep{deepfri}, recovered catalytic residues with \emph{post-hoc}
gradient saliency, which explains each prediction separately, one protein at a time. \tool{}
instead learns its substructures jointly across proteins, and they transfer to new proteins such as
the held-out enzymes.

\vspace{-1ex}
\subsection{\tool{} adapts to the natural granularity of functional units}
\label{sec:results-granularity}
The granularity of the blobs is set by three hyperparameters. The number of blobs $K$
sets how many substructures \tool{} forms, the seed radius $r$ sets how far each blob can extend
from its seed, and the Hoyer weight $\lambda_{\text{H}}$ sets how many residues within that radius
a blob keeps. Functional units range from catalytic sites of a few residues to structural domains
spanning hundreds, so the granularity that suits prediction may differ across tasks. We first vary
$K$ and $r$ together on three ProteinShake tasks with different targets, then vary
$\lambda_{\text{H}}$ on \texttt{SCOP-FAM}.

\begin{figure}[ht]
    \centering
    \includegraphics[width=1\linewidth]{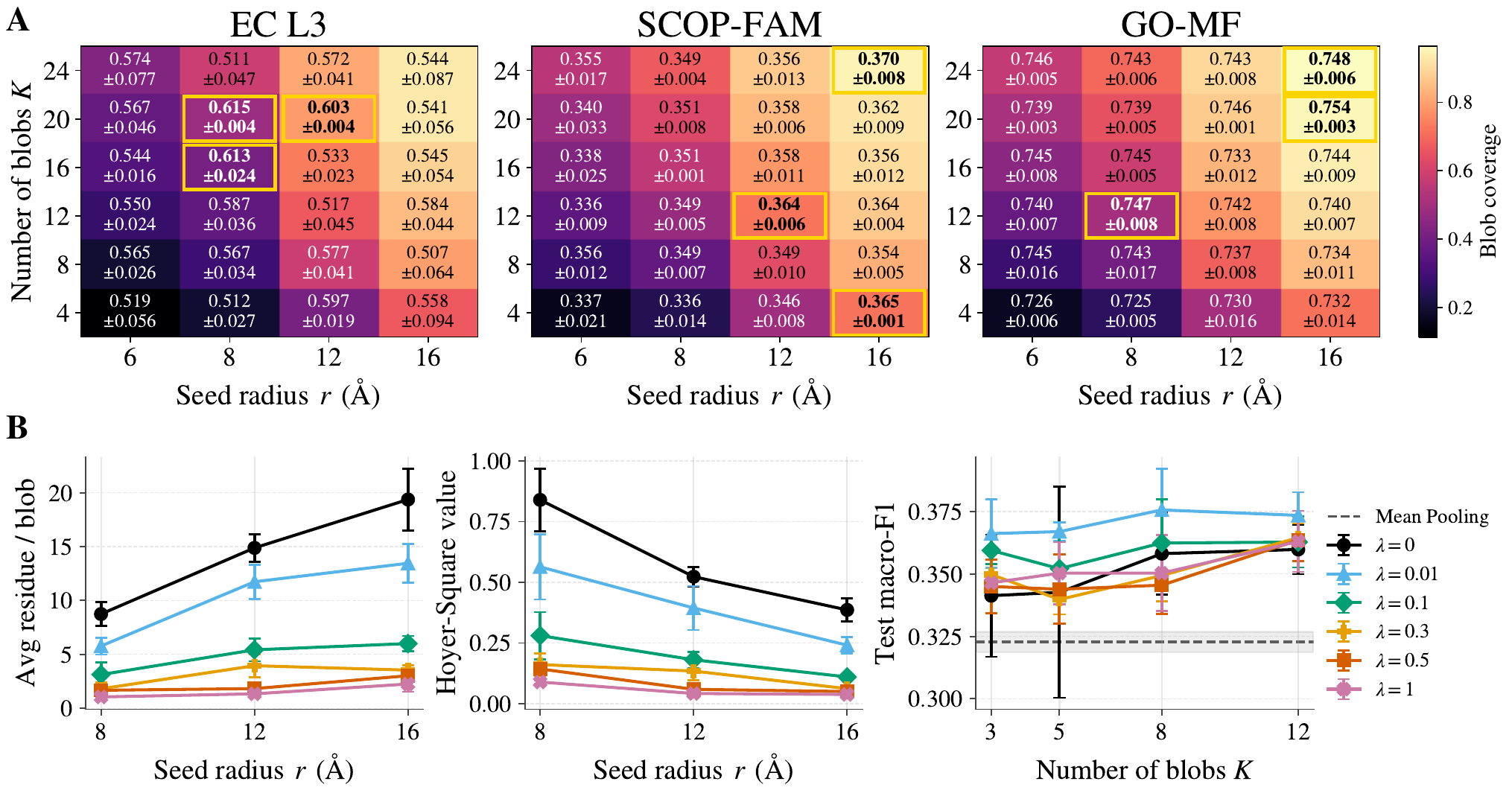}
\caption{\textbf{\tool{} hyperparameter ablations.}
\textbf{A.} Test metric over the number of blobs $K$ and seed radius $r$ on three ProteinShake tasks, with ESM2-T30 on the sequence split and $\lambda_{\text{H}}=0.3$. Color shows blob coverage, the fraction of residues within $r$ of any selected seed averaged over test proteins, and boxes mark the three best configurations of each task. \textbf{B.} Blob size, Hoyer-Square value and macro-F1 on \texttt{SCOP-FAM} as $\lambda_{\text{H}}$ varies.}
\vspace{-2ex}
    \label{fig:abalation}
\end{figure}

\vspace{-1ex}
\paragraph{Number of blobs and seed radius.}
Whether prediction improves with more blobs or a larger seed radius depends on the task, as
Figure~\ref{fig:abalation}A shows for ESM2-T30 on the sequence split. \texttt{SCOP-FAM} classifies
entire structural domains spanning hundreds of residues, and its test metric rises with
blob coverage and peaks at $0.370$ in the high-coverage corner at $K=24, r=16$. \texttt{EC-L3}
shows no such trend, and \tool{} reaches its best test metric of $0.615$ at $K=20, r=8$,
where blob coverage is about half. \texttt{GO-MF} gains little from coverage and
depends more on the number of blobs. Many configurations lie within seed variation of each task's
best, so the trends are more informative than individual optima. These trends follow the spatial scale
of each task's labels. Structural domains span most of a chain, whereas enzyme classes
depend on localized catalytic sites.

\vspace{-1ex}
\paragraph{Hoyer regularization weight.}
Increasing $\lambda_{\text{H}}$ compacts blobs by an order of magnitude at nearly no cost in
accuracy. Figure~\ref{fig:abalation}B tracks blob size, the Hoyer-Square value and macro-F1 on
\texttt{SCOP-FAM}. At $K{=}5, r{=}12$\,\AA{}, mean blob size at a membership threshold of
$0.5$ falls from about $15$ residues at $\lambda_{\text{H}}{=}0$ to $1.3$ at $\lambda_{\text{H}}{=}1$,
and the Hoyer-Square value drops from $0.52$ to $0.04$. At $r{=}16$\,\AA{}, blob size falls
from about $19$ to $2.3$ residues. Across $K\in\{3,5,8,12\}$ at $r{=}12$\,\AA{}, macro-F1 for
$\lambda_{\text{H}} \in (0, 1]$ overlaps with $\lambda_{\text{H}}{=}0$ within seed variation.
Every tested regularized configuration also exceeds the mean-pooling baseline. The prediction
therefore needs only a few residue embeddings per blob.

\Needspace{17\baselineskip}
\vspace{-1ex}
\subsection{\tool{} is computationally efficient and scalable}
\label{sec:results-cost}
\begin{wraptable}{r}{0.5\textwidth}
\vspace{-\intextsep}
\setlength{\abovecaptionskip}{0pt}
\centering
\small
\setlength{\tabcolsep}{4.5pt}
\caption{\textbf{Training time and memory.}}
\label{tab:training_time}
\vspace{0.4em}
\resizebox{\linewidth}{!}{%
\begingroup\color{black}\renewcommand{\crchange}[1]{#1}%
\begin{tabular}{lccc}
\toprule
Configuration & s/epoch & \crchange{Time} $\times$ & \crchange{Memory} $\times$ \\
\midrule
\mean                              & $2.98$ & $1.00$ & $1.00$ \\
\midrule
\multicolumn{4}{l}{\tool, sweeping $K$ at $r=12$} \\
\quad $K=4$              & $3.86$ & $1.29$ & $3.43$ \\
\quad $K=8$              & $3.86$ & $1.30$ & $3.45$ \\
\quad $K=12$             & $3.85$ & $1.29$ & $3.46$ \\
\quad $K=16$             & $4.21$ & $1.41$ & $3.48$ \\
\quad $K=20$             & $4.46$ & $1.50$ & $3.49$ \\
\quad $K=24$             & $4.24$ & $1.42$ & $3.51$ \\
\midrule
\multicolumn{4}{l}{\tool, sweeping $r$ at $K=12$} \\
\quad $r=6$              & $3.93$ & $1.32$ & $3.46$ \\
\quad $r=8$              & $3.78$ & $1.27$ & $3.46$ \\
\quad $r=16$             & $3.70$ & $1.24$ & $3.46$ \\
\quad $r=20$             & $3.91$ & $1.31$ & $3.46$ \\
\bottomrule
\end{tabular}\endgroup}
\end{wraptable}

The \tool{} neural blob partitioner adds only modest training time, and its cost scales linearly with protein length. On VenusX \texttt{Act} with ESM2-T30 residue embeddings (Table~\ref{tab:training_time}), \tool{} at $K{=}12, r{=}12$ runs at $1.29\times$ the per-epoch wall-clock of a mean-pooling MLP whose classifier head is matched to the \tool{} MIL classifier, and uses $3.46\times$ its peak GPU memory. The seed radius $r$ is essentially free: both time and memory remain flat across all tested radii. Cost grows gently with the number of blobs $K$, with wall-clock at most $1.50\times$ and memory growing by under $3\%$ from $K{=}4$ to $K{=}24$. These trends are consistent with the partitioner's complexity. Every partitioner operation is asymptotically linear in the sequence length $N$: seed scoring costs $O(N D d_h)$, the seed-query and node-key projections are $O((N+K) D d_a)$, scaled-dot-product seed-to-residue attention is $O(K N d_a)$, and the soft-membership-weighted pool is $O(K N D)$, where $d_h = d_a = 128$ are the seed-MLP and attention hidden widths. Total cost per batch is therefore $O(B N D (d_h + d_a + K))$, which does not depend on $r$ and is dominated by the $d_h + d_a$ constants since $K \ll d_h + d_a$.
\vspace{-1ex}
\section{Related Work}
\vspace{-1ex}
\label{sec:related_work}
\paragraph{Protein representations for function prediction.}
Function prediction builds on learned protein representations \citep{prl}, from sequence
language models \citep{esm}, structure-aware language models \citep{su_saprot_2023} and structure
encoders \citep{gearnet,pst}. These models describe a protein either residue by residue or as a
single vector. The vector comes from mean or attention pooling \citep{stark_light_2021,ilse2018attention},
or from graph pooling that merges residues into learned clusters \citep{ying2018hierarchical,poolreview}.
Intermediate units are usually fixed in advance, such as curated or predicted domains
\citep{scope,cath,merizo}, surface patches \citep{gainza2020deciphering} or extracted subgraphs \citep{subgnn}.
Functional sites fall between these scales. A single vector dilutes a site of a few residues, and
pooled clusters or relaxed graph partitions \citep{erdos,pnc} need not be spatially compact.
\tool{} learns this intermediate unit for each task, as compact and spatially local blobs on top of
any residue encoder, and predicts from the blobs alone.
\vspace{-1ex}
\paragraph{Localizing protein function without residue labels.}
Supervised predictors localize sites from residue or surface annotations
\citep{scannet,masif}, but such labels are scarce and each model usually targets one type of
site. PARSE avoids training by matching local structure embeddings to annotated sites \citep{russ}, but it still needs those annotations. Localization can also be learned from global labels, as with class activation maps for images
\citep{zhou2016learning,selvaraju2017grad} and attention-based multiple-instance learning for labeled
bags \citep{ilse2018attention}. For proteins, gradient-based maps on function classifiers \citep{deepfri}
and the attention weights of pooling models highlight individual residues. These per-residue scores
show a site only after thresholding, and nothing makes the top residues form one
unit. \tool{} localizes through its blobs instead, so the evidence for each prediction is a small
set of cohesive candidate sites. To our knowledge, \tool{} is the first method to recover
experimentally-curated catalytic sites by \emph{directly mining over learned substructures}.
\vspace{-1ex}
\paragraph{Explaining predictions with substructures.}
Explanations of protein models are usually computed after training. Integrated gradients
\citep{sundararajan2017axiomatic} and gradient-based maps \citep{deepfri} score each residue for one
prediction, and sparse autoencoders find language-model features that track binding sites or motifs
\citep{sae,simon_interplm_2025}. Because they are computed apart from the decision, two such
explanations can disagree. \tool{}
makes the explanation part of the prediction. Its MIL predictor sees only the blobs, so the
highest-attention blobs are the substructures the prediction rests on.

\vspace{-1ex}
\section{Limitations and Future Work}
\label{sec:limitations}
\vspace{-1ex}

\tool{} learns which substructures are predictive of function across many proteins, but it does not establish which substructure performs which function in an individual protein. A protein that both binds a ligand and catalyzes a reaction, for example, receives a single set of importance scores rather than one set per function. Confirming the role of each region in a given protein still requires targeted experiments such as mutagenesis. Finally, \tool{} still needs protein-level function labels for training, so it can only discover substructures tied to the functions it was trained to predict. Such labels are sparse and concentrated in well-studied protein families. Discovering functional substructures without function labels, directly from sequences and structures, would extend the search to functions that have not yet been characterized.

\vspace{-1ex}
\section{Conclusion}
\label{sec:conclusion}
\vspace{-1ex}

\tool connects protein function prediction with functional substructure discovery. Compatible with any protein encoder, it predicts function from a small set of learned, spatially local blobs, so that each prediction can be traced to specific substructures. Across diverse protein functions, such as reaction catalysis and domain families, it is competitive with strong baselines using both sequence-only and sequence--structure encoders. Trained only on protein-level labels, it localizes many types of annotated functional sites, including ligand-binding interfaces and catalytic residues. These results support functional substructure discovery across the unannotated proteome.

\begin{ack}
We thank Asif Solanki for technical support and the Vanderbilt Center for Structural Biology for hosting our compute infrastructure. Xin Wang acknowledges fellowship support from the Computational Biology and Biomedical Informatics (CBB) program at Yale University.
\end{ack}

\bibliographystyle{plainnat}
\bibliography{citations}

\clearpage
\appendix
\etocdepthtag.toc{appendix}

\begingroup
\renewcommand{\textcolor}[2]{#2}
\hypersetup{hidelinks,linktoc=all}
\etocsettagdepth{main}{none}
\etocsettagdepth{appendix}{subsection}
\etocsettagdepth{checklist}{none}
\etocsetnexttocdepth{subsection}
\etocsettocstyle{\section*{Appendix contents}}{}
\tableofcontents
\endgroup
\clearpage
\raggedbottom 

\section{Code Availability and Reproducibility}
\label{appendix:code}
The code, the trained M-CSA model, and instructions for reproducing all experiments are publicly available at \url{https://github.com/OliverLaboratory/BioBlobs}. An interactive viewer at \url{https://oliverlaboratory.com/resources/bioblobs/} lets readers compare the model's blobs with the annotated catalytic residues of M-CSA enzymes.

\section{\texorpdfstring{\tool{} Pipeline Details}{BioBlobs Pipeline Details}}
\label{appendix:pipeline}
Figure~\ref{fig:pipeline_detailed} expands the overview in Figure~\ref{fig:bioblobs_pipeline} into the steps of Section~\ref{sec:method}, and Algorithm~\ref{alg:bioblobs} gives the corresponding forward pass and training objective. Structure-aware encoders also read structure tokens (dashed arrow), and the structure provides the C$_\alpha$ coordinates $\mathbf{X} \in \mathbb{R}^{N \times 3}$ used by the partitioner.

\paragraph{Neural blob partitioner.}
The straight-through estimator passes gradients through the discrete choice of each seed by combining the one-hot selection $\mathbf{e}^{(k)}$ with the softmax $\mathbf{w}^{(k)}$:
\begin{equation}
    \mathbf{e}^{(k)} = \mathrm{one\_hot}\!\left(\arg\max_i\; w_i^{(k)}\right), \qquad \tilde{\mathbf{w}}^{(k)} = \mathbf{e}^{(k)} + \left(\mathbf{w}^{(k)} - \mathrm{sg}(\mathbf{w}^{(k)})\right),
    \label{eq:seed_st}
\end{equation}
where $\mathrm{sg}(\cdot)$ denotes the stop-gradient operator, and the seed embedding is $\mathbf{z}_{\text{seed}}^{(k)} = \sum_i \tilde{w}_i^{(k)} \mathbf{z}_i$. The temperature is annealed across epochs as $\tau_{\text{seed}} = \max(\tau_{\min},\, \tau_{\text{init}} \cdot \gamma^{\text{epoch}})$, progressively sharpening the softmax, and $\tau_{\text{mem}}$ follows the same schedule.

\paragraph{Residue scores.}
The M-CSA results in Figure~\ref{fig:csa_recovery} score residues by their maximum membership over blobs, and the VenusX results in Table~\ref{tab:venusx} score them by their attention-weighted membership.

\begin{figure}[ht]
    \centering
\definecolor{bbEnc}{HTML}{87CEEB}
\definecolor{bbPart}{HTML}{F4A582}
\definecolor{bbMil}{HTML}{8FD18F}
\definecolor{bbRvq}{HTML}{B39DDB}
\begin{tikzpicture}[
  font=\scriptsize,
  >={Stealth[length=1.8mm]},
  comp/.style={draw, rounded corners=2pt, fill=white, align=center,
    text width=19mm, minimum height=10mm, inner sep=2pt, line width=0.7pt},
  io/.style={draw=black!55, rounded corners=2pt, fill=black!4, align=center,
    text width=19mm, minimum height=8mm, inner sep=2pt},
  loss/.style={draw=black!55, dashed, rounded corners=2pt, fill=white, align=center,
    text width=19mm, minimum height=8mm, inner sep=2pt},
  tag/.style={font=\scriptsize, inner sep=1pt},
  arr/.style={->, line width=0.6pt},
  hop/.style={preaction={draw=white, line width=3pt, -}},
  frame/.style={rounded corners=4pt, line width=0.8pt, inner sep=0pt},
  ftitle/.style={anchor=north west, font=\footnotesize\bfseries, inner sep=0pt},
]
\node[io] (seq) at (1.05,0) {Sequence};
\node[comp, draw=bbEnc!80!black, fill=bbEnc!35] (enc) at (3.8,0) {Protein encoder};
\node[io] (str) at (6.55,0) {Structure};
\draw[arr] (seq) -- (enc);
\draw[arr, dashed] (str) -- (enc);

\begin{scope}[on background layer]
  \draw[frame, draw=bbPart!85!black, fill=bbPart!12] (-0.2,-0.8) rectangle (13.3,-4.3);
\end{scope}
\node[ftitle] at (0.0,-0.95) {Neural blob partitioner};
\node[comp, draw=bbPart!85!black] (score) at (1.05,-2.25) {Seed scoring};
\node[comp, draw=bbPart!85!black] (select) at (3.8,-2.25) {Sequential seed selection ($K$ rounds, straight-through)};
\node[comp, draw=bbPart!85!black] (cand) at (6.55,-2.25) {Candidate set within radius $r$};
\node[comp, draw=bbPart!85!black] (memb) at (9.3,-2.25) {Soft membership (attention and proximity bias)};
\node[comp, draw=bbPart!85!black] (pool) at (12.05,-2.25) {Blob pooling (membership-weighted mean)};
\node[loss] (hoyer) at (9.3,-3.65) {Hoyer-Square penalty $\mathcal{L}_{\text{Hoyer}}$};

\draw[arr] (score) -- node[tag, above] {$s_i$} (select);
\draw[arr] (select) -- node[tag, above, font=\tiny] {seeds} (cand);
\draw[arr] (cand) -- node[tag, above] {$\mathcal{C}^{(k)}$} (memb);
\draw[arr] (memb) -- node[tag, above] {$\mathbf{M}$} (pool);
\draw[arr] (memb) -- (hoyer);
\node[anchor=north west, align=left, font=\scriptsize, inner sep=0pt] at (0.05,-3.1) {%
  $\mathcal{C}^{(k)} = \{\, i : \|\mathbf{x}_i - \mathbf{x}_{\text{seed}}^{(k)}\|_2 \le r \,\}$\\[1pt]
  $m_{ki} = \sigma\big((a_{ki} + \lambda_\rho\, \rho_{ki}) / \tau_{\text{mem}}\big), \ i \in \mathcal{C}^{(k)}$\\[1pt]
  $\mathbf{b}_k = \sum_i m_{ki}\, \mathbf{z}_i \,/\, \sum_i m_{ki}$};

\draw[line width=0.6pt] (enc.south) -- (3.8,-1.35);
\draw[line width=0.6pt] (1.05,-1.35) -- (12.05,-1.35);
\foreach \x in {1.05,3.8,9.65,12.05} {
  \fill (\x,-1.35) circle (1.1pt);
  \draw[arr] (\x,-1.35) -- (\x,-1.75);
}
\node[tag, right] at (3.8,-0.62) {$\mathbf{Z}$};
\draw[arr, hop] (str.south) -- (cand.north);
\node[tag, left] at (6.55,-0.66) {$\mathbf{X}$};
\fill (6.55,-0.52) circle (1.1pt);
\draw[arr, hop] (6.55,-0.52) -- (8.95,-0.52) -- (8.95,-1.75);

\draw[arr] (pool.south) -- node[tag, right] {$\mathbf{b}_k$} (12.05,-4.6) -- (1.05,-4.6) -- (1.05,-6.1);

\begin{scope}[on background layer]
  \draw[frame, draw=bbMil!80!black, fill=bbMil!14] (2.75,-4.95) rectangle (10.55,-7.95);
\end{scope}
\node[ftitle] at (2.95,-5.1) {MIL predictor};
\node[comp, dashed, draw=bbRvq!90!black, fill=bbRvq!12] (rvq) at (1.05,-6.6) {Residual vector quantization (optional)};
\node[loss] (commit) at (1.05,-8.15) {Commitment penalty};
\node[comp, draw=bbMil!80!black] (inst) at (4.1,-5.95) {Feature transform};
\node[comp, draw=bbMil!80!black] (gate) at (4.1,-7.25) {Attention gate};
\node[comp, draw=bbMil!80!black] (agg) at (6.85,-6.6) {Attention-weighted pooling};
\node[comp, draw=bbMil!80!black] (head) at (9.3,-6.6) {MLP\\prediction head};
\node[io] (pred) at (12.05,-6.6) {Protein-level prediction};
\node[loss] (task) at (12.05,-8.15) {Task loss $\mathcal{L}_{\text{task}}$};

\draw[arr] (rvq) -- (commit);
\draw[line width=0.6pt] (rvq.east) -- (2.6,-6.6);
\fill (2.6,-6.6) circle (1.1pt);
\draw[arr] (2.6,-6.6) |- (inst.west);
\node[tag, above] at (2.31,-6.6) {$\widehat{\mathbf{b}}_k$};
\draw[arr] (2.6,-6.6) |- (gate.west);
\draw[arr] (inst.east) -- node[tag, above] {$\mathbf{u}_k$} ++(0.4,0) |- ([yshift=2mm]agg.west);
\draw[arr] (gate.east) -- node[tag, below] {$\alpha_k$} ++(0.4,0) |- ([yshift=-2mm]agg.west);
\draw[arr] (agg) -- node[tag, above] {$\mathbf{z}_{\text{bag}}$} (head);
\draw[arr] (head) -- (pred);
\draw[arr] (pred) -- (task);
\node[anchor=north west, align=left, font=\scriptsize, inner sep=0pt] at (5.95,-7.22) {%
  $\alpha_k = \mathrm{softmax}_k\big(\mathbf{w}_a^\top \mathbf{b}_k + c_a\big)$\\[1pt]
  $\mathbf{z}_{\text{bag}} = \sum_k \alpha_k\, \mathbf{u}_k$};

\node[io, text width=36mm] (interp) at (5.0,-8.85) {Blob importances $\alpha_k$ and residue scores from $\mathbf{M}$ (and $\alpha_k$)};
\draw[arr] (gate.south) -- (gate.south |- interp.north);
\end{tikzpicture}
    \caption{Detailed \tool{} pipeline.}
    \label{fig:pipeline_detailed}
\end{figure}

\begin{algorithm}[H]
\caption{\tool{} forward pass and training objective for one protein.}
\label{alg:bioblobs}
\small
\begin{algorithmic}[1]
\Require Residue embeddings $\mathbf{Z} = [\mathbf{z}_1, \dots, \mathbf{z}_N]$, C$_\alpha$ coordinates $\mathbf{X}$, number of blobs $K$, radius $r$, temperatures $\tau_{\text{seed}}, \tau_{\text{mem}}$, weights $\lambda_\rho, \lambda_{\text{H}}$
\Ensure Protein-level prediction, membership matrix $\mathbf{M}$, blob importances $\boldsymbol{\alpha}$, training loss $\mathcal{L}$
\Statex \textit{Seed selection}
\State \label{alg:l:seed}$s_i \gets f_{\text{seed}}(\mathbf{z}_i)$ for every residue $i$; \; $\mathcal{U} \gets \{1, \dots, N\}$ \Comment{unselected residues}
\For{$k = 1, \dots, K$}
    \State $\mathbf{w}^{(k)} \gets \mathrm{softmax}(\tilde{\mathbf{s}}^{(k)} / \tau_{\text{seed}})$ \Comment{$\tilde{\mathbf{s}}^{(k)}$: scores masked to $\mathcal{U}$}
    \State $v_{\text{seed}}^{(k)} \gets \arg\max_i w_i^{(k)}$; \; $\mathbf{e}^{(k)} \gets \mathrm{one\_hot}(v_{\text{seed}}^{(k)})$
    \State $\tilde{\mathbf{w}}^{(k)} \gets \mathbf{e}^{(k)} + \big(\mathbf{w}^{(k)} - \mathrm{sg}(\mathbf{w}^{(k)})\big)$ \Comment{straight-through, Eq.~\eqref{eq:seed_st}}
    \State $\mathbf{z}_{\text{seed}}^{(k)} \gets \sum_i \tilde{w}_i^{(k)} \mathbf{z}_i$; \; $\mathbf{x}_{\text{seed}}^{(k)} \gets \mathbf{x}_{v_{\text{seed}}^{(k)}}$; \; $\mathcal{U} \gets \mathcal{U} \setminus \{v_{\text{seed}}^{(k)}\}$
\EndFor
\Statex \textit{Blob expansion and pooling}
\For{$k = 1, \dots, K$}
    \State $\mathcal{C}^{(k)} \gets \{i : \|\mathbf{x}_i - \mathbf{x}_{\text{seed}}^{(k)}\|_2 \le r\}$ \Comment{candidates within radius $r$}
    \State $m_{ki} \gets \sigma\big((a_{ki} + \lambda_\rho \rho_{ki}) / \tau_{\text{mem}}\big)$ for $i \in \mathcal{C}^{(k)}$, else $m_{ki} \gets 0$ \Comment{Eqs.~\eqref{eq:attn_prox}--\eqref{eq:membership}}
    \State \label{alg:l:clamp}$m_{k, v_{\text{seed}}^{(k)}} \gets 1$ \Comment{seed belongs to its blob}
    \State \label{alg:l:pool}$\mathbf{b}_k \gets \sum_i m_{ki} \mathbf{z}_i \,/\, \big(\sum_i m_{ki} + \varepsilon\big)$ \Comment{Eq.~\eqref{eq:blob_embedding}}
\EndFor
\If{RVQ is enabled}
    \State replace each $\mathbf{b}_k$ by its quantized embedding $\widehat{\mathbf{b}}_k$ \Comment{Eq.~\eqref{eq:rvq}}
\EndIf
\Statex \textit{MIL prediction}
\State $\mathbf{u}_k \gets \mathrm{ReLU}(\mathbf{W}_\phi \mathbf{b}_k + \mathbf{c}_\phi)$ for every blob $k$
\State $\boldsymbol{\alpha} \gets$ softmax over valid blobs of $\mathbf{w}_a^\top \mathbf{b}_k + c_a$ \Comment{Eq.~\eqref{eq:mil_attention}}
\State \label{alg:l:pred}$\mathbf{z}_{\text{bag}} \gets \sum_k \alpha_k \mathbf{u}_k$; \; prediction $\gets \mathrm{MLP}(\mathbf{z}_{\text{bag}})$
\Statex \textit{Training objective}
\State \label{alg:l:loss}$\mathcal{L} \gets \mathcal{L}_{\text{task}} + \lambda_{\text{H}} \mathcal{L}_{\text{Hoyer}}$, plus the RVQ commitment penalty if enabled \Comment{Eq.~\eqref{eq:objective}}
\State \Return prediction, $\mathbf{M}$, $\boldsymbol{\alpha}$, $\mathcal{L}$
\end{algorithmic}
\end{algorithm}

\clearpage
\section{Math Notation}
\label{appendix:notation}

This appendix provides a reference for the mathematical notation used in the paper. Table~\ref{tab:notation} lists the key symbols introduced in each part of the model and in the localization evaluation.

{\small
\setlength{\LTcapwidth}{\linewidth}
\begingroup\color{black}\renewcommand{\crchange}[1]{#1}%
\begin{longtable}{@{}p{0.28\linewidth}>{\raggedright\arraybackslash}p{0.68\linewidth}@{}}
\caption{\textbf{Math notation.} Symbols are grouped by the part of the paper that defines them.}
\label{tab:notation}\\
\toprule
\textbf{Symbol} & \textbf{Meaning} \\
\midrule
\endfirsthead
\toprule
\textbf{Symbol} & \textbf{Meaning} \\
\midrule
\endhead
\midrule
\multicolumn{2}{r@{}}{\footnotesize\textit{Continued on the next page}} \\
\endfoot
\bottomrule
\endlastfoot
\rowcolor{black!7}[0pt][0pt]\multicolumn{2}{@{}l@{}}{\textbf{Protein representation} \quad {\footnotesize Section~\ref{sec:encoder}}} \\*
\addlinespace[2pt]
$N$ & Number of valid residues in a protein \\*
$\mathcal{A}$ & Standard 20-letter amino acid alphabet \\*
$\mathbf{X} \in \mathbb{R}^{N \times 3}$ & Matrix of C$_\alpha$ coordinates. Rows are $\mathbf{x}_i$ \\*
$\mathbf{Z} \in \mathbb{R}^{N \times D}$ & Residue-level embeddings from the protein encoder. Rows are $\mathbf{z}_i$ \\*
$D$ & Encoder embedding dimension \\
\addlinespace[4pt]
\rowcolor{black!7}[0pt][0pt]\multicolumn{2}{@{}l@{}}{\textbf{Seed selection} \quad {\footnotesize Section~\ref{sec:partitioner}}} \\*
\addlinespace[2pt]
$K$ & Number of blobs proposed per protein \\*
$s_i = f_{\text{seed}}(\mathbf{z}_i)$ & Seed score of residue $i$ from the seed scoring MLP $f_{\text{seed}}$ \\*
$\mathbf{w}^{(k)},\ \tau_{\text{seed}}$ & Seed softmax over unselected residues at step $k$, and its temperature \\*
$v_{\text{seed}}^{(k)},\ \mathbf{z}_{\text{seed}}^{(k)},\ \mathbf{x}_{\text{seed}}^{(k)}$ & Residue index, embedding and C$_\alpha$ coordinate of the $k$-th seed \\
\addlinespace[4pt]
\rowcolor{black!7}[0pt][0pt]\multicolumn{2}{@{}l@{}}{\textbf{Blob expansion} \quad {\footnotesize Section~\ref{sec:partitioner}}} \\*
\addlinespace[2pt]
$r$ & Seed radius (\AA) bounding the candidate set of each blob \\*
$\mathcal{C}^{(k)},\ n_k$ & Candidate residues within radius $r$ of seed $k$, and their number $n_k = |\mathcal{C}^{(k)}|$ \\*
$a_{ki},\ \rho_{ki}$ & Attention score and proximity bias of residue $i$ for seed $k$ \\*
$\lambda_\rho$ & Weight on the proximity bias \\*
$\ell_{ki}$ & Membership logit of residue $i$ for seed $k$, $\ell_{ki} = a_{ki} + \lambda_\rho\, \rho_{ki}$ \\*
$m_{ki},\ \tau_{\text{mem}}$ & Soft membership of residue $i$ in blob $k$, and the membership temperature \\*
$\mathbf{M} \in [0,1]^{K \times N}$ & Membership matrix. Rows are blobs, and a residue can belong to more than one blob \\*
$\mathbf{m}_k$ & Membership vector of blob $k$ over its candidates \\
\addlinespace[4pt]
\rowcolor{black!7}[0pt][0pt]\multicolumn{2}{@{}l@{}}{\textbf{MIL predictor} \quad {\footnotesize Section~\ref{sec:mil}}} \\*
\addlinespace[2pt]
$\mathbf{b}_k \in \mathbb{R}^D$ & Membership-weighted mean of residue embeddings for blob $k$ \\*
$\mathbf{B} \in \mathbb{R}^{K \times D}$ & Matrix of blob embeddings \\*
$\mathbf{u}_k$ & Transformed blob embedding (output of the feature transform) \\*
$\alpha_k$ & MIL attention weight on blob $k$, used as its importance score \\*
$\mathbf{z}_{\text{bag}}$ & Protein-level representation, $\mathbf{z}_{\text{bag}} = \sum_k \alpha_k\, \mathbf{u}_k$ \\
\addlinespace[4pt]
\rowcolor{black!7}[0pt][0pt]\multicolumn{2}{@{}l@{}}{\textbf{Residual vector quantization} \quad {\footnotesize Section~\ref{sec:mil}}} \\*
\addlinespace[2pt]
$Q,\ V$ & Number of RVQ stages, and of code vectors per codebook \\*
$\boldsymbol{\mu}_{q,j},\ j_{k,q}$ & Code vector $j$ of stage $q$, and the code index of blob $k$ at stage $q$ \\*
$\widehat{\mathbf{b}}_k \in \mathbb{R}^{D}$ & Quantized blob embedding, replacing $\mathbf{b}_k$ when RVQ is enabled \\
\addlinespace[4pt]
\rowcolor{black!7}[0pt][0pt]\multicolumn{2}{@{}l@{}}{\textbf{Training objectives} \quad {\footnotesize Section~\ref{sec:objectives}}} \\*
\addlinespace[2pt]
$\mathcal{L},\ \mathcal{L}_{\text{task}}$ & Training objective, and the task loss (cross-entropy or binary cross-entropy) \\*
$\mathrm{HS}_k \in [1/n_k,\, 1]$ & Hoyer-Square value of blob $k$, lower for sparser memberships \\*
$\mathcal{L}_{\text{Hoyer}}$ & Hoyer-Square penalty, the average of $\mathrm{HS}_k$ over valid blobs \\*
$\lambda_{\text{H}},\ \lambda_{\mathrm{RVQ}}$ & Weights of the Hoyer-Square and RVQ commitment penalties \\
\addlinespace[4pt]
\rowcolor{black!7}[0pt][0pt]\multicolumn{2}{@{}l@{}}{\textbf{Seed alignment and localization evaluation} \quad {\footnotesize Appendix~\ref{appendix:localization}}} \\*
\addlinespace[2pt]
$\mathcal{L}_{\text{seed-align}}$ & Seed-level alignment loss, used only by the \tool{} (+ seed align) models in Table~\ref{tab:venusx} \\*
$c_i$ & Localization score of residue $i$. For \tool{}, $\max_k m_{ki}$ (Figure~\ref{fig:csa_recovery}) or $\sum_k \alpha_k m_{ki}$ (Table~\ref{tab:venusx}) \\*
$y_i \in \{0,1\}$ & Residue annotation used for evaluation \\*
$L = \sum_i y_i$ & Number of annotated residues in a protein \\*
$E_p$ & AUPRC enrichment of protein $p$, its AUPRC divided by $L/N$ \\
\end{longtable}
\endgroup
}

\clearpage

\section{Dataset Details}
\label{appendix:datasets}
Table~\ref{tab:dataset_info} summarizes the datasets, and Section~\ref{sec:dataset} describes their splits. ProteinShake and VenusX evaluate protein function prediction. VenusX, whose site annotations come from InterPro~\citep{blum_interpro_2025}, and M-CSA also evaluate functional site localization. The model evaluated on M-CSA is trained on SwissProt enzymes.

\begin{table}[ht]
\centering
\small
\caption{Datasets used in our experiments. SwissProt trains the model evaluated on M-CSA, and its counts cover the training and validation proteins of up to 1{,}000 residues. M-CSA counts all evaluated enzymes, 74 of which form the held-out set.}
\label{tab:dataset_info}
\setlength{\tabcolsep}{4pt}
\newcommand{\dstwoline}[2]{\begin{tabular}[t]{@{}l@{}}#1\\#2\end{tabular}}
\begingroup\color{black}\renewcommand{\crchange}[1]{#1}%
\begin{tabularx}{\columnwidth}{@{}l r r >{\raggedright\arraybackslash}X l l@{}}
\toprule
\textbf{Dataset} & \textbf{Proteins} & \textbf{Classes} & \textbf{Functional aspect} & \textbf{Task type} & \textbf{Metric} \\
\midrule
\multicolumn{6}{@{}l}{\textbf{ProteinShake}} \\
\quad EC-L3   & 15{,}057  & 31    & Reaction catalysis (sub-subclass)   & Multi-class & \crchange{Macro-F1} \\
\quad SCOP-FAM & 10{,}066  & 3{,}042 & Structural family                 & Multi-class & \crchange{Macro-F1} \\
\quad SCOP-SF & 10{,}066  & 1{,}537 & Structural superfamily            & Multi-class & \crchange{Macro-F1} \\
\quad Pfam    & 31{,}109  & 5{,}163 & Sequence/domain family            & Multi-class & \crchange{Macro-F1} \\
\quad GO-MF   & 32{,}633  & 5{,}127 & Molecular function                & Multi-label & F$_{\max}$ \\
\addlinespace[2pt]
\multicolumn{6}{@{}l}{\textbf{VenusX}} \\
\quad Act    & 1{,}860   & 132 & Catalytic active sites               & \dstwoline{Multi-class}{Localization} & \dstwoline{Macro-F1}{AUROC, AUPRC, F1@$L$} \\
\quad BindI  & 2{,}050   & 76  & Ligand binding interfaces            & \dstwoline{Multi-class}{Localization} & \dstwoline{Macro-F1}{AUROC, AUPRC, F1@$L$} \\
\quad Evo    & 12{,}975  & 739 & Evolutionarily conserved sites       & \dstwoline{Multi-class}{Localization} & \dstwoline{Macro-F1}{AUROC, AUPRC, F1@$L$} \\
\quad Motif  & 2{,}508   & 289 & Conserved sequence/structure motifs  & \dstwoline{Multi-class}{Localization} & \dstwoline{Macro-F1}{AUROC, AUPRC, F1@$L$} \\
\addlinespace[2pt]
\multicolumn{6}{@{}l}{\textbf{M-CSA evaluation}} \\
\quad SwissProt & 203{,}782 & 242 & Reaction catalysis (sub-subclass) & Multi-class & -- \\
\quad M-CSA & 883 & -- & Catalytic residues & Localization & AUPRC enrichment \\
\bottomrule
\end{tabularx}\endgroup
\end{table}

\section{Implementation Details}
\label{appendix:implementation}
\paragraph{Function-prediction baselines and metrics.}
We use frozen residue embeddings from ESM2-T30~\citep{esm}, a sequence-only transformer trained with masked language modeling, and SaProt-650M~\citep{su_saprot_2023}, a sequence-structure transformer trained on residues augmented with Foldseek structural tokens. Mean pooling averages residue embeddings into a fixed-size protein representation. Attention pooling instead scores each residue with a single linear projection, softmax-normalizes the scores across valid residues and takes the attention-weighted sum. The graph baselines are GCN, GIN and GAT, and the graph-pooling baselines are DiffPool, TopK and SAGPool. All baselines and \tool{} share the same MLP classifier architecture. The graph and graph-pooling baselines are tuned on validation data, with eight trials on EC-L3 and SCOP-FAM and two on the other tasks. Macro-F1 averages per-class F1 over the classes present in the test labels or predictions, and $F_{\max}$ for GO-MF is the best F1 over 21 score thresholds.

\paragraph{\tool{} hyperparameters.}
\label{implementation}
Table~\ref{tab:hyperparameters} lists the \tool{} hyperparameter space. Inter-blob self-attention is an optional self-attention layer over the transformed blob embeddings $\mathbf{u}_k$ before they are pooled. We report mean $\pm$ sample standard deviation over three seeds at the selected configuration. The M-CSA model uses $K=8$, $r=12$\,\AA{}, $\lambda_{\text{H}}=0.5$ and $\lambda_\rho=0.25$, with batch size 512 for 100 epochs, and the VenusX settings follow below.

\begin{table}[ht]
\centering
\small
\caption{\textbf{\tool{} hyperparameters.} Searched hyperparameters take values from the listed sets, and the other settings are fixed.}
\label{tab:hyperparameters}
\begin{tabular}{@{}ll@{}}
\toprule
\textbf{Hyperparameter} & \textbf{Value} \\
\midrule
\multicolumn{2}{@{}l}{\textit{Searched}} \\
\quad Number of blobs $K$ & $\{4, 8, 12, 16, 20, 24, 32\}$ \\
\quad Seed radius $r$ & $\{6, 8, 12, 16, 20, 24\}$\,\AA \\
\quad Hoyer-Square weight $\lambda_{\text{H}}$ & $\{10^{-4}, 3{\times}10^{-4}, 10^{-3}, 3{\times}10^{-3}, 10^{-2}, 3{\times}10^{-2}, 0.1, 0.3\}$ \\
\quad Inter-blob self-attention & Off or on \\
\addlinespace[2pt]
\multicolumn{2}{@{}l}{\textit{Partitioner}} \\
\quad Seed-scoring MLP hidden width $d_h$ & 128 \\
\quad Attention projection width $d_a$ & 128 \\
\quad Proximity weight $\lambda_\rho$ & 0.5 \\
\quad Temperatures $\tau_{\text{seed}}$ and $\tau_{\text{mem}}$ & $\tau_{\text{init}} = 1.0$, decay $\gamma = 0.95$ per epoch, $\tau_{\min} = 0.25$ \\
\addlinespace[2pt]
\multicolumn{2}{@{}l}{\textit{MIL predictor}} \\
\quad Projection width & 512 \\
\quad MLP head hidden widths & 2{,}048 and 1{,}024 \\
\quad Dropout & 0.1 \\
\addlinespace[2pt]
\multicolumn{2}{@{}l}{\textit{Training}} \\
\quad Optimizer & Adam, learning rate $10^{-3}$, no weight decay \\
\quad Learning-rate schedule & Cosine, 5 warmup epochs \\
\quad Batch size & 128 \\
\quad Epochs & 60 or 100 \\
\quad Early stopping & Validation primary metric, patience 15 \\
\bottomrule
\end{tabular}
\end{table}

\paragraph{VenusX.}
Without residue supervision, \tool{} fixes $K=20$, $r=16$\,\AA{} and learning rate $10^{-4}$, without RVQ, and searches $\lambda_{\text{H}}\in\{0,\,3\times10^{-4},\,10^{-3},\,3\times10^{-3},\,0.01,\,0.1\}$ and dropout in $\{0.1,0.3\}$. Training uses Adam with weight decay $10^{-4}$, batch size 32, a cosine schedule over at most 100 epochs and early-stopping patience 30. Mean and attention pooling use one fixed configuration, with dropout 0.1, learning rate $3\times10^{-4}$, weight decay $10^{-4}$, batch size 32 and at most 60 epochs with patience 15. The \tool{} (+ seed align) models use seed-site supervision and annotation-informed granularity, so they are not a controlled ablation of supervision. We keep the supplied MF50 splits, which are not protein-disjoint, and all methods share the same 186, 203, 1{,}291 and 249 test proteins for \texttt{Act}, \texttt{BindI}, \texttt{Evo} and \texttt{Motif}.

\paragraph{Hardware.}
The original experiments used a single node equipped with 8\,$\times$\,NVIDIA H100 80\,GB HBM3 GPUs, two AMD EPYC 9454 48-core CPUs (96 physical cores / 192 threads), and 1.5\,TB of system RAM. The VenusX runs without residue supervision use NVIDIA RTX 5090 GPUs.

\clearpage
\section{\texorpdfstring{$K,r$}{K,r} Grid Search on ProteinShake}

\begin{figure}[ht]
    \centering
    \includegraphics[width=1\linewidth]{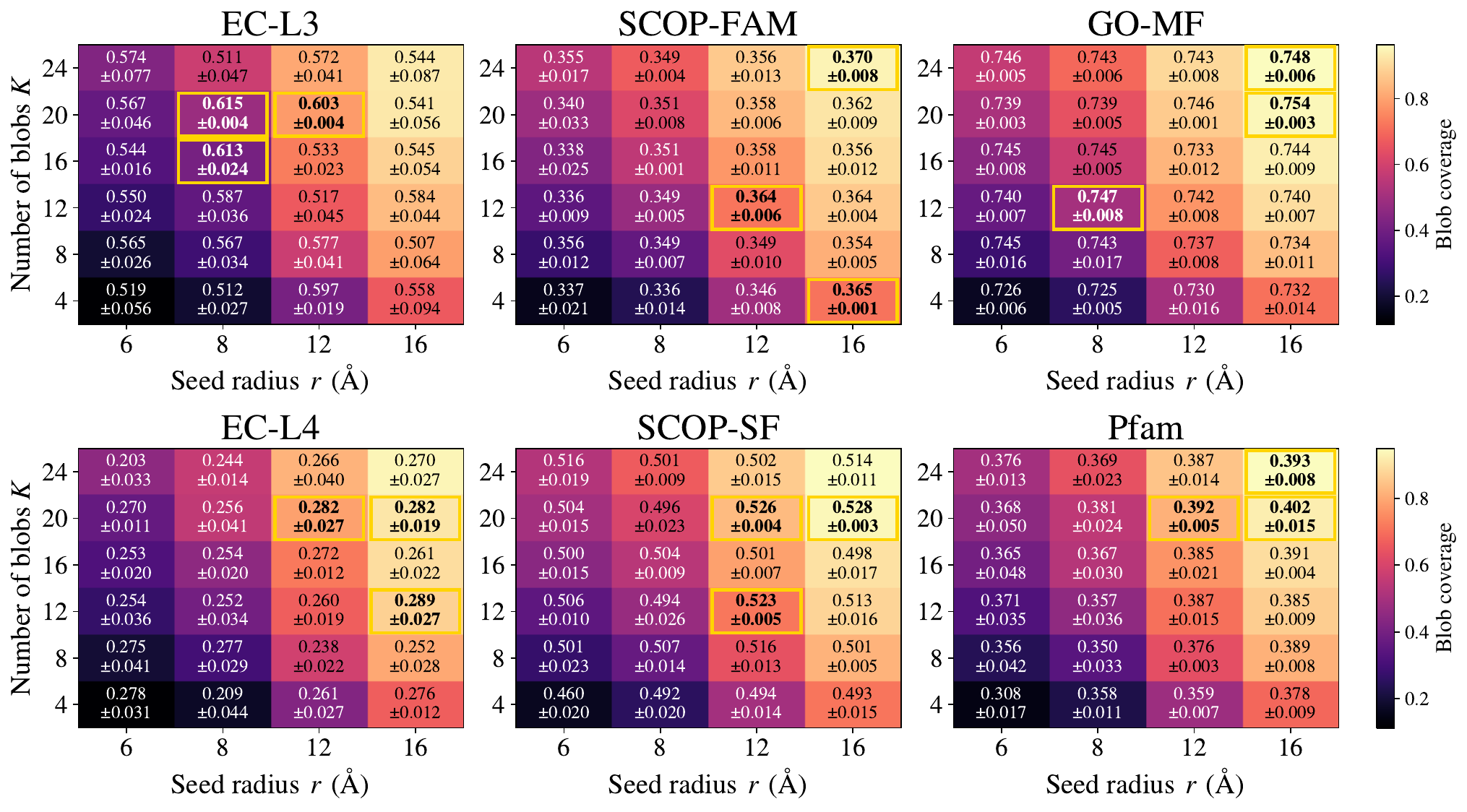}
    \caption{$K,r$ grid search on ProteinShake. Color denotes mean blob coverage, the fraction of residues within radius $r$ of any selected seed.}
    \label{fig:kr_grid_search_all}
\end{figure}

Blob coverage tracks prediction most closely on tasks whose labels describe large regions. Figure~\ref{fig:kr_grid_search_all} shows the test metric over the full $K,r$ grid, and Table~\ref{tab:kr_trends} correlates it with blob coverage across the 24 configurations. The correlation lies between 0.71 and 0.92 on \texttt{Pfam} and \texttt{SCOP-FAM} in every setting, whereas \texttt{EC-L3} and \texttt{GO-MF} have the two lowest correlations for each encoder and split. This follows the spatial scale of the labels discussed in Section~\ref{sec:results-granularity}.

\begin{table}[ht]
\centering
\caption{\textbf{Prediction versus blob coverage on ProteinShake.} Spearman
correlation of the test metric with blob coverage across the 24 $(K, r)$ configurations, for each encoder and split. Each configuration's
metric and coverage are averaged over three seeds. On validation metrics, \texttt{EC-L3} and
\texttt{GO-MF} also have the lowest coverage correlations in every setting.}
\label{tab:kr_trends}
\small
\setlength{\tabcolsep}{4pt}
\begingroup\color{black}\renewcommand{\crchange}[1]{#1}%
\begin{tabular}{lcccc}
\toprule
 & \multicolumn{2}{c}{ESM2-T30} & \multicolumn{2}{c}{SaProt-650M} \\
\cmidrule(lr){2-3}\cmidrule(lr){4-5}
Task & Seq. & Struct. & Seq. & Struct. \\
\midrule
\texttt{EC-L3} & $-0.03$ & $0.05$ & $0.41$ & $0.57$ \\
\texttt{GO-MF} & $0.35$ & $0.32$ & $0.64$ & $0.54$ \\
\texttt{Pfam} & $0.92$ & $0.71$ & $0.91$ & $0.74$ \\
\texttt{SCOP-FAM} & $0.81$ & $0.71$ & $0.91$ & $0.91$ \\
\texttt{SCOP-SF} & $0.46$ & $0.51$ & $0.96$ & $0.95$ \\
\bottomrule
\end{tabular}\endgroup
\end{table}

\section{Residual Vector Quantization}
\label{appendix:rvq}
\paragraph{Method.}
Equation~\eqref{eq:rvq} in Section~\ref{sec:method} defines the quantization.
We update each codebook with exponential moving averages (decay 0.99) of its
assigned residuals and pass gradients to the blob embeddings through a
straight-through estimator~\citep{vqvae}. We also minimize the commitment penalty
$\lambda_{\mathrm{RVQ}}\,|\mathcal{V}_{\mathrm{blob}}|^{-1}
\sum_{k\in\mathcal{V}_{\mathrm{blob}}}
\|\boldsymbol{\delta}_{k,0}-\mathrm{sg}(\widehat{\mathbf{b}}_k)\|_2^2$,
which keeps blob embeddings close to their quantized reconstructions.

\paragraph{Setup.}
The RVQ model is the M-CSA model, with frozen ESM2-T30 embeddings, $Q=8$ stages and $V=1{,}024$ code vectors per stage, and the continuous model is the same model without RVQ.
Both checkpoints are selected by EC validation macro-F1 and scored by the
attention-weighted membership $c_i=\sum_k\alpha_k m_{ki}$ on the full cached
M-CSA set (874 evaluations per model), without the training-homology exclusion
used for Table~\ref{tab:mcsa_localization}.

\begin{table}[ht]
\centering
\caption{\textbf{Effect of RVQ on catalytic residue localization on M-CSA.}
Protein-level medians for \tool{} with and without RVQ on the full cached
evaluation set ($n=874$ per model).}
\label{tab:rvq_localization_ablation}
\small
\begingroup\color{black}\renewcommand{\crchange}[1]{#1}%
\begin{tabular}{lccc}
\toprule
Configuration & AUROC & AUPRC & F1@$L$ \\
\midrule
\tool{} without RVQ & 0.911 & 0.184 & 0.167 \\
\tool{} with RVQ & 0.942 & 0.331 & 0.333 \\
\bottomrule
\end{tabular}\endgroup
\end{table}

Table~\ref{tab:rvq_localization_ablation} shows that RVQ raises the median AUPRC from 0.184 to 0.331. Each row is a single checkpoint
evaluated on a different set from Table~\ref{tab:mcsa_localization}, so the
comparison measures neither variation across seeds nor the gain on the
74-protein set.
\section{Functional Site Localization}
\label{appendix:localization}

\subsection{Evaluation}
\label{appendix:localization_metrics}
For one protein, let $c_i$ be the localization score of residue $i$ and
$y_i\in\{0,1\}$ its annotation, with $N$ evaluated residues and
$L=\sum_{i=1}^{N}y_i$ positives. We use each dataset's residue masks and exclude
proteins without both positive and negative residues. For \tool{}, $c_i$ is the
attention-weighted membership $\sum_k\alpha_k m_{ki}$, except in
Figure~\ref{fig:csa_recovery}A and Table~\ref{tab:mcsa_localization}, which use
the maximum membership $\max_k m_{ki}$.

\paragraph{Metrics.}
AUROC is the probability that a positive residue outranks a negative one, with
ties counted as half. AUPRC is average precision, without trapezoidal interpolation. F1@$L$
takes the $L$ highest-scoring residues, so its precision, recall and F1 all equal
$H/L$, where $H$ is the number of positives among them. For M-CSA, the AUPRC enrichment
of protein $p$ is $E_p=\mathrm{AUPRC}_p/(L_p/N_p)$, and we report the median of
$E_p$ over proteins rather than a ratio of medians.

\paragraph{Aggregation and uncertainty.}
For VenusX, we take the protein median within each seed and report the mean and
standard deviation over three seeds. For M-CSA, we report medians over the 74
held-out enzymes, with 95\% percentile intervals from 5{,}000 protein-bootstrap
resamples shared across methods. These intervals reflect protein sampling for
fixed trained models, not training variation.

\paragraph{VenusX protein function prediction.}
For \tool{} without residue supervision, model selection uses validation macro-F1
over the fixed class vocabulary, counting classes absent from both labels
and predictions as zero, so the denominator stays constant during hyperparameter
search, early stopping and checkpoint selection.
Table~\ref{tab:venusx_classification} reports test macro-F1 as mean and standard deviation over the checkpoints of Table~\ref{tab:venusx}. Without residue supervision, \tool{} has the highest mean on \texttt{Act}, \texttt{BindI} and \texttt{Evo} with ESM2-T30 and on \texttt{Act} and \texttt{Evo} with SaProt-650M, and attention pooling has the highest mean on the other datasets.
\begin{table}[ht]
\centering
\caption{\textbf{Protein function prediction on VenusX.}}
\label{tab:venusx_classification}
\small
\setlength{\tabcolsep}{4pt}
\begin{tabular}{llcccc}
\toprule
 & & \multicolumn{4}{c}{Test macro-F1 $\uparrow$} \\
\cmidrule(lr){3-6}
Encoder & Method & \texttt{Act} & \texttt{BindI} & \texttt{Evo} & \texttt{Motif} \\
\midrule
\multirow{4}{*}{ESM2-T30}
 & Mean pooling & $0.539_{\pm0.026}$ & $0.824_{\pm0.004}$ & $0.749_{\pm0.005}$ & $0.385_{\pm0.022}$ \\
 & Attention pooling & $0.633_{\pm0.031}$ & $0.879_{\pm0.020}$ & $0.778_{\pm0.015}$ & $\mathbf{0.460_{\pm0.018}}$ \\
 & \tool{} & $\mathbf{0.653_{\pm0.021}}$ & $\mathbf{0.924_{\pm0.030}}$ & $\mathbf{0.798_{\pm0.009}}$ & $0.442_{\pm0.029}$ \\
 & \tool{} (+ seed align) & $0.621_{\pm0.012}$ & $0.918_{\pm0.009}$ & $0.792_{\pm0.010}$ & $0.416_{\pm0.025}$ \\
\midrule
\multirow{4}{*}{SaProt-650M}
 & Mean pooling & $0.553_{\pm0.016}$ & $0.828_{\pm0.039}$ & $0.792_{\pm0.009}$ & $0.454_{\pm0.038}$ \\
 & Attention pooling & $0.624_{\pm0.002}$ & $\mathbf{0.878_{\pm0.032}}$ & $0.820_{\pm0.002}$ & $\mathbf{0.470_{\pm0.048}}$ \\
 & \tool{} & $\mathbf{0.667_{\pm0.046}}$ & $0.871_{\pm0.040}$ & $\mathbf{0.823_{\pm0.013}}$ & $0.416_{\pm0.026}$ \\
 & \tool{} (+ seed align) & $0.613_{\pm0.026}$ & $0.911_{\pm0.048}$ & $0.823_{\pm0.007}$ & $0.423_{\pm0.006}$ \\
\bottomrule
\end{tabular}
\end{table}

\subsection{Seed alignment}
\label{appendix:alignment}
\label{appendix:seed_align}
Seed alignment treats the softmax of the seed logits over valid residues as a categorical distribution over candidate seed positions and maximizes the probability mass on annotated residues $\mathbf{y} \in \{0,1\}^N$:
\begin{equation}
p_i \;=\; \frac{\exp(s_i)}{\sum_{j=1}^{N} \exp(s_j)},
\qquad
\mathcal{L}_{\text{seed-align}} \;=\; -\log\!\Bigl(\,\sum_{i=1}^{N} y_i\, p_i\Bigr).
\end{equation}
This is the negative log-likelihood that a seed sampled from $p$ falls within an annotated site. It acts on the seed scores and reaches the memberships only through the seed choice. The loss is applied only to proteins with at least one annotated residue. The \tool{} (+ seed align) models in Table~\ref{tab:venusx} use it, and the other \tool{} models do not.

\subsection{VenusX baselines}
\label{appendix:localization_baselines}
The learned baselines use frozen ESM2-T30 or SaProt-650M residue embeddings and are trained
to predict protein-level labels. Table~\ref{tab:venusx} scores residues with IG on mean pooling and with the attention weights of attention pooling.

\paragraph{Integrated gradients (IG).}
Using integrated gradients~\citep{sundararajan2017axiomatic}, we attribute the
predicted-class logit to the input residue embeddings of
the mean-pooling classifier. Let $\mathbf{Z}$ be the
embedding matrix, $f_c(\mathbf{Z})$ the logit for target class $c$, and $\mathbf{Z}^{0}$ a reference matrix.
The attribution of embedding coordinate $d$ at residue $i$ is
\begin{equation}
\mathrm{IG}_{id}(\mathbf{Z})
=(Z_{id}-Z^0_{id})\int_0^1
\left.\frac{\partial f_c(\mathbf{U})}{\partial U_{id}}
\right|_{\mathbf{U}=\mathbf{Z}^0+t(\mathbf{Z}-\mathbf{Z}^0)}\,\mathrm{d}t.
\end{equation}
We use a zero reference and approximate the integral with 16 equally spaced
right-endpoint samples. The class predicted for the original input is fixed
along the integration path. Residue scores are the signed sums
$c_i=\sum_d\mathrm{IG}_{id}$. No absolute-value transformation is applied.
The IG rows reuse the trained classifier without further optimization.

\paragraph{Residue burial and uniform random scores.}
Residue burial counts C$_\alpha$ atoms within 10\,\AA{} of each residue,
so densely packed residues receive higher scores. The random baseline assigns independent uniform scores
to valid residues, with reproducible random seeds.

\subsection{Blob overlap with annotated sites on VenusX}
\label{appendix:venusx_best_blob}
For the \tool{} (+ seed align) model, a blob's soft recall on an annotated site is its membership mass on annotated residues divided by the site size, and its soft precision divides that mass by its total membership mass. Table~\ref{tab:venusx_best_blob} maximizes recall, precision and F1 separately over blobs and averages these site-level maxima within each protein and then across proteins, so the three maxima can come from different blobs. This analysis uses the annotations to find the best overlap and does not evaluate a blob selected by the model's attention weights.
\begin{table}[ht]
\centering
\caption{\textbf{Blob overlap with annotated sites on VenusX.} Single-seed analysis of the \tool{} (+ seed align) model in Table~\ref{tab:venusx} with frozen ESM2-T30 embeddings. Site size is the median annotated-site size in residues, and blob size is the mean of within-protein median sizes at membership threshold $0.5$.}
\label{tab:venusx_best_blob}
\small
\setlength{\tabcolsep}{6pt}
\begingroup\color{black}\renewcommand{\crchange}[1]{#1}%
\begin{tabular}{lccccc}
\toprule
Dataset & Site size & Blob size & Max. recall & Max. precision & Max. F1 \\
\midrule
Act & 12 & 8.0 & 0.482 & 0.753 & 0.529 \\
BindI & 13 & 17.5 & 0.671 & 0.618 & 0.574 \\
Evo & 15 & 17.0 & 0.581 & 0.602 & 0.513 \\
Motif & 42 & 14.8 & 0.358 & 0.864 & 0.451 \\
\bottomrule
\end{tabular}\endgroup
\end{table}

\subsection{M-CSA baselines}
The published enzyme-function models retain their released weights and
architectures. The three attribution methods instead use our
EC-trained ESM2-T30 mean-pooling classifier. All rows use the common
74-protein set described in Section~\ref{sec:csa_recovery}.
Of the 79 held-out enzymes, 74 have scores from every method.
The homology exclusion is relative to our EC training set. It does not
establish exclusion from the published models' training sets.

\begin{table}[ht]
\centering
\caption{\textbf{Full M-CSA localization results.}
All 14 methods on the 74 held-out enzymes of Figure~\ref{fig:csa_recovery}A, as
protein-level medians. Brackets give 95\% protein-bootstrap confidence intervals for
AUPRC enrichment. The linear probe is trained on catalytic-residue labels and serves
as a supervised ceiling.}
\label{tab:mcsa_localization}
\small
\setlength{\tabcolsep}{5pt}
\begingroup\color{black}\renewcommand{\crchange}[1]{#1}%
\begin{tabular}{lccc}
\toprule
Method & AUPRC enrichment [95\% CI] & AUPRC & AUROC \\
\midrule
\tool{} & $10.7\times$ [8.5, 12.1] & $0.159$ & $0.913$ \\
\midrule
SAE (summed magnitude) & $6.2\times$ [4.6, 8.7] & $0.114$ & $0.790$ \\
SAE (latent count) & $5.0\times$ [3.6, 7.3] & $0.085$ & $0.764$ \\
ESM2 attention & $4.2\times$ [3.1, 5.8] & $0.067$ & $0.753$ \\
\midrule
ProtDETR (encoder self-attention) & $3.2\times$ [2.4, 4.1] & $0.053$ & $0.704$ \\
ProtDETR (decoder cross-attention) & $1.8\times$ [1.3, 2.5] & $0.029$ & $0.576$ \\
ProteInfer (CAM) & $1.5\times$ [1.3, 1.7] & $0.027$ & $0.529$ \\
DeepFRI (Grad-CAM) & $1.4\times$ [1.2, 1.9] & $0.025$ & $0.522$ \\
\midrule
Mean pooling (IG) & $1.8\times$ [1.4, 2.2] & $0.031$ & $0.545$ \\
Mean pooling (gradient $\times$ input) & $1.7\times$ [1.3, 2.4] & $0.031$ & $0.547$ \\
Mean pooling (Grad-CAM) & $1.4\times$ [1.0, 3.3] & $0.029$ & $0.510$ \\
\midrule
Burial ($-$rSASA) & $1.7\times$ [1.5, 2.0] & $0.025$ & $0.646$ \\
Uniform random & $1.4\times$ [1.2, 1.5] & $0.021$ & $0.500$ \\
\midrule
Linear probe (catalytic labels) & $28.5\times$ [23.9, 34.7] & $0.573$ & $0.978$ \\
\bottomrule
\end{tabular}\endgroup
\end{table}

\paragraph{ProtDETR.}
ProtDETR~\citep{yang2025interpretable} combines ESM-1b residue embeddings with a transformer and learned
functional queries for enzyme classification. We evaluate two attention
readouts from the released model. The encoder score averages the
self-attention received by each residue over query positions and encoder
layers. The decoder score averages query-to-residue cross-attention over
decoder layers for the functional query with the largest non-null EC logit.
Both readouts use head-averaged attention weights. Inputs are truncated to
1,022 residues to respect the encoder's sequence-length limit.

\paragraph{ProteInfer (CAM).}
ProteInfer~\citep{sanderson_proteinfer_2023} is a dilated convolutional network for protein function prediction.
Class activation mapping~\citep{zhou2016learning} scores each residue by the inner product between
its final convolutional representation and the output weights for the target
EC class. We use the annotated EC class when represented in the model's
vocabulary, otherwise an available class sharing its first three EC levels,
and finally the highest-scoring predicted class if neither is available.
The scores retain their sign.

\paragraph{DeepFRI (Grad-CAM).}
DeepFRI~\citep{deepfri} combines sequence features with graph convolutions on
a residue-contact graph. We use its released EC model, with C$_\alpha$ contacts
defined by a 10\,\AA{} cutoff. Grad-CAM~\citep{selvaraju2017grad} weights each channel of the concatenated
graph-convolution features by its target-score gradient averaged over residues,
then sums the weighted channels at each residue. We use the annotated EC
class, its three-level parent when available, or the highest-scoring prediction.
The released implementation rescales the resulting scores to $[0,1]$ without
rectifying them. Guided-gradient filtering is disabled.

\paragraph{Integrated gradients, gradient $\times$ input and Grad-CAM.}
IG and gradient $\times$ input attribute an EC-class logit of our mean-pooling classifier to
its frozen residue embeddings. The annotated EC-L3 class is used when present
in the classifier vocabulary. Otherwise the predicted class is used.
IG~\citep{sundararajan2017axiomatic} uses a zero embedding reference with 32 integration steps.
Here, residue scores are absolute values after summing over embedding
coordinates: $c_i=|\sum_d\mathrm{IG}_{id}|$ for IG and
$c_i=|\sum_d Z_{id}\,\partial f_c(\mathbf{Z})/\partial Z_{id}|$ for
gradient $\times$ input~\citep{ancona2017towards}.
Grad-CAM~\citep{selvaraju2017grad} on the same classifier weights each residue embedding by
the residue-averaged gradient of the target logit and rectifies the result. The
classifier mean-pools its inputs, so this equals gradient $\times$ input with negative
contributions set to zero.

\paragraph{ESM2 attention.}
This baseline uses no labels. It scores each residue by the attention it receives in
ESM2-T30~\citep{esm}, averaged over all 30 layers and 20 heads.

\paragraph{Sparse autoencoder (SAE).}
A BatchTopK sparse autoencoder~\citep{bussmann2024batchtopk} with 5{,}120 latents and $k=32$
is trained without labels on the final-layer ESM2-T30 embeddings that \tool{} also uses,
from 4{,}000 EC training proteins that share no 30\%-identity cluster with an M-CSA enzyme.
At inference, one protein's residues form a single batch. The magnitude score sums each residue's retained activations,
and the count score counts its retained latents.

\paragraph{Burial ($-$rSASA).}
This label-free heuristic scores each residue by its negative relative solvent accessibility,
computed with the Shrake--Rupley algorithm~\citep{shrake_environment_1973} on the deposited
structure and normalized by theoretical maximum accessibilities~\citep{tien_maximum_2013}.
Bound ligands count as occluders, so the score partly reflects where ligands bind.

\paragraph{Linear probe.}
As a supervised ceiling, an L2-regularized logistic regression with balanced class weights
predicts catalytic residues from layer-29 ESM2-T30 embeddings. Under
cross-validation grouped by 30\%-identity cluster, each enzyme is scored by a model that
never saw its cluster.

\paragraph{Uniform random scores.}
We assign independent uniform scores to residues using a fixed random seed.
Finite-sample average precision need not equal the positive-residue prevalence, so the table reports the measured random baseline.

\clearpage

\end{document}